\documentclass[%
reprint,
amsmath,
amssymb,
aps,
prc,
floatfix,
]{revtex4-2}
\usepackage{graphicx}% Include figure files
\usepackage{epstopdf}
\usepackage{dcolumn}% Align table columns on decimal point
\usepackage{bm}% bold math
\usepackage[colorlinks=true, urlcolor=blue, linkcolor=blue,citecolor=blue]{hyperref}% add hypertext capabilities
\usepackage[mathlines]{lineno}% Enable numbering of text and display math
\usepackage{longtable}
\usepackage{amsmath}
\usepackage{natbib}
\usepackage{float}
\usepackage{booktabs}
\usepackage{etoolbox}
\usepackage{blindtext}
\usepackage{gensymb}
\usepackage{placeins}
\usepackage{xcolor}
\usepackage[T1]{fontenc}
\usepackage[utf8]{inputenc}
\makeatletter
\newif\ifLT@nocaption
\preto\longtable{\LT@nocaptiontrue}
\appto\endlongtable{%
\ifLT@nocaption
\addtocounter{table}{\m@ne}%
\fi}
\preto\LT@caption{%
\noalign{\global\LT@nocaptionfalse}}
\makeatother
\begin{document}
\preprint{APS/123-QED}

\title{Probability for synthesis of superheavy nuclei with $\textbf{Z=121}$}% Force line breaks with \\
%\thanks{A footnote to the article title}%

\author{R. Zargini}
\email{reza.zargini@pnu.ac.ir,reza.zargini@gmail.com}
\affiliation{%
 Department of Physics, Payame Noor University (PNU), P.O. Box 19395-4697, Tehran, Iran}%
 %\altaffiliation[Also at ]{Physics Department, Payame Noor University.}%Lines break automatically or can be forced with \\
\author{S. A. Seyyedi}%
\email{a.seyyedi@pnu.ac.ir
}
\affiliation{%
Department of Physics, Payame Noor University (PNU), P.O. Box 19395-4697, Tehran, Iran}%

%\collaboration{}%\noaffiliation

%\author{Charlie Author}
 %\homepage{http://www.Second.institution.edu/~Charlie.Author}
%\affiliation{
 %Second institution and/or address\\
 %This line break forced% with \\
%}%
%\affiliation{
%Third institution, the second for Charlie Author
%}%
%\author{Delta Author}
%\affiliation{%
 %Authors' institution and/or address\\
 %This line break forced with \textbackslash\textbackslash
%}%

%\collaboration{CLEO Collaboration}%\noaffiliation

\date{\today}% It is always \today, today,
             %  but any date may be explicitly specified
\begin{abstract}
In this study the empirical method \cite{RN464} is used for calculating the evaporation residue (ER) cross section in the synthesis of superheavy nuclei (SHN). The superheavy nuclei examined in this work fall within the range $Z=112-118$. The theoretical calculations show good agreement with the experimental data. Furthermore, this model, along with an investigation of the optimal incident energy (OIE), is used to calculate the ER cross section for a hypothetical heavy system with $Z=121$. Five promising combinations are suggested for synthesis of SHN with $Z=121$: (1) ${^{50}}\mathrm{Ti}+{^{252}}\mathrm{Es}$, with the maximum ER cross section $\sigma_{3n}=24.5~\mathrm{fb}$, at the optimal incident energy, $\mathrm{OIE}=230~\mathrm{MeV}$; (2) ${^{50}}\mathrm{Ti}+{^{254}}\mathrm{Es}$, with the maximum ER cross section $\sigma_{3n}=11.8~\mathrm{fb}$, at the optimal incident energy, $\mathrm{OIE}=229~\mathrm{MeV}$; (3) ${^{51}}\mathrm{V}+{^{251}}\mathrm{Cf}$, with the maximum ER cross section $\sigma_{3n}=1.2~\mathrm{fb}$, at the optimal incident energy, $\mathrm{OIE}=238~\mathrm{MeV}$; (4) ${^{51}}\mathrm{V}+{^{249}}\mathrm{Cf}$, with the maximum ER cross section $\sigma_{3n}=1.0~\mathrm{fb}$, at the optimal incident energy, $\mathrm{OIE}=238~\mathrm{MeV}$; and (5) ${^{54}}\mathrm{Cr}+{^{247}}\mathrm{Bk}$, with the maximum ER cross section $\sigma_{2n}=0.9~\mathrm{fb}$, at the optimal incident energy, $\mathrm{OIE}=243~\mathrm{MeV}$.
\end{abstract}
\maketitle

\section{Introduction}

Low energy fusion reactions with energy $E\le15~\mathrm {MeV}/\mathrm {nucleon}$ are important tools for the synthesis of superheavy nuclei (SHN). Theoretical studies could provide extremely valuable insights into the optimization of expensive experimental efforts and pave the roads to the synthesis of SHN \cite{RN455,RN456,RN457,RN458,RN459,RN460,RN461,RN532,RN528,RN462,RN464,RN564,RN465,RN466,RN541,RN467,RN525,RN468,RN469,RN470,RN471,RN521,RN562,RN563}. Supported by theoretical studies and employing heavy ion accelerators, nuclei with $Z<112$ have been produced via cold fusion reactions using double magic target ${^{208}}\mathrm{Pb}$. In contrast, nuclei with $Z=112-118$ were generated via hot fusion reactions and employing double magic projectile ${^{48}}\mathrm{Ca}$ and actinide targets \cite{RN472,RN473,RN474,RN475,RN476,RN477,RN478,RN479,RN480,RN481,RN482,RN483,RN484,RN485,RN486,RN487,RN565,RN566}. Different models have been utilized in studies of the probability of synthesis of SHN. The models include the empirical method \cite{RN464,RN564}, the dinuclear system (DNS) model \cite{RN455,RN563,RN456,RN458}, the two-step model \cite{RN488}, the fusion by diffusion model \cite{RN489}, and the nuclear collectivization model \cite{RN490}. Calculating total potential is crucial in the empirical model, for which different models, such as double folding, Woods-Saxon, Skyrme energy density formalism, and proximity, have been developed~\cite{RN496,RN499}. To calculate nuclear potential, considering nuclei deformation is important \cite{RN458,RN461,RN467}. Our group employed nuclei deformations along with the significance of surface energy coefficients in the investigation of ER cross sections \cite{RN491}. There are three fundamental stages in elucidating SHN synthesis with the empirical model \cite{RN466}: (1) The projectile and target overcome the potential barrier by the system center of mass energy and stick to each other. Wong, Glas, Mosel, and others developed models to calculate the capture cross section $\sigma_\mathrm{Capture}$. \cite{RN494,RN554,RN460}. (2)Projectile and target nuclei combine and form the excited compound nucleus, so calculating the formation of the compound nucleus probability $P_{CN}$ is significant \cite{RN464,RN528,RN555,RN354}. (3)After forming the compound nucleus, it should reach an equilibrium state and survive against fission and other decay modes. The survival probability $W_\mathrm{sur}$ is one of the most important parameters that should be calculated; different models obtain this term \cite{RN465}. Scientists are interested in achieving the island of stability on the top of the nuclear chart, expanding the nuclear chart, and producing heavier nuclei \cite{RN532}. As mentioned above, oganeson, with $Z=118$, is the heaviest nucleus in the periodic table of elements. In the previous study, our group attempted to calculate the probability synthesis of superheavy nuclei with $Z=119$ and $120$ \cite{RN491}. In this work we present the probability synthesis of superheavy nuclei with $Z=121$. To synthesize superheavy nuclei with $Z=121$, targets heavier than californium ($\mathrm{Cf}$) are required, and producing targets heavier than $\mathrm{Cf}$ is not a trivial task. For instance, einsteinium (${^{254}}\mathrm{Es}$) can be synthesized in microgram scales, approximately three orders of magnitude less than typically required values for SHN synthesis \cite{RN493}. Hence, one could consider projectiles heavier than ${^{48}}\mathrm{Ca}$, \textit{i.e.},${^{50}}\mathrm{Ti}$, ${^{49-50}}\mathrm{V}$, ${^{52-54}}\mathrm{Cr}$, ${^{45-55}}\mathrm{Mn}$, ${^{54-58}}\mathrm{Fe}$, ${^{59}}\mathrm{Co}$, and ${^{64}}\mathrm{Ni}$ and actinide targets such as ${^{235-237}}\mathrm{Np}$, ${^{238-242,244}}\mathrm{Pu}$, ${^{241,243}}\mathrm{Am}$, ${^{243-248}}\mathrm{Cm}$, ${^{247,249}}\mathrm{Bk}$, ${^{248-252}}\mathrm{Cf}$, and ${^{252,254}}\mathrm{Es}$. Several theoretical investigations have been performed on the probability of synthesis of SHN with $Z=121$. The most suggested combinations were ${^{50}}\mathrm{Ti}+{^{252,254}}\mathrm{Es}$,  ${^{54}}\mathrm{Cr}+{^{249}}\mathrm{Bk}$, ${^{50}}\mathrm{V}+{^{251}}\mathrm{Cf}$, and ${^{64}}\mathrm{Ni}+{^{235}}\mathrm{Np}$. The calculated ER cross section range was from $1.6~\mathrm{pb}$ to $0.03803~\mathrm{fb}$ \cite{RN461,RN462,RN467,RN468,RN469,RN470,RN471,RN552,RN553}.\\ 
In this manuscript the model and theory are explained in sec.~\ref{Section2}. The results on the comparison with experimental data for superheavy nuclei in the range of $Z=112-118$ are described in sec.~\ref{Section3-1}. Finally, employing the obtained model, our results of the probability of the synthesis of superheavy nuclei with $Z=121$ will be discussed in sec.~\ref{Section3-2}.

\section{\label{Section2}Model and theory}

\subsection{The capture cross section}

The barrier penetration model developed by Wong has been widely used to describe the fusion reactions at the energies close to, or greater than that of, the barrier \cite{RN494}. The capture cross section is expressed as the sum of the cross sections for each partial wave $l$ \cite{RN455}:
\begin{eqnarray}
\sigma_\mathrm{cap}=\frac{\pi\hbar^2}{2\mu E_{\mathrm{c.m.}}}\sum_{l=0}^{l_{max}}{(2l+1)}T_l\left(E_{\mathrm{c.m.}},l\right),\label{eq:1}
\end{eqnarray}
where $\mu$ is the reduced mass of the interacting nuclei, $E_{\mathrm{c.m.}}$ is the center of mass of the colliding systems, and $l$ is the angular momentum. $T_l\left(E_{\mathrm{c.m.}},l\right)$ denotes the penetration probability of the potential barrier for the $l\mathrm{th}$ partial wave and calculated according to the Hill-Wheeler equation \cite{RN495}: 
\begin{eqnarray}
T_l\left(E_{\mathrm{c.m.}},l\right)=\left\{1+\mathrm{exp}\left[\frac{2\pi\left(E_l-E_{\mathrm{c.m.}}\right)}{\hbar\omega_l}\right]\right\}^{-1}.\label{eq:2}
\end{eqnarray}
The Wong formula is used to calculate the capture cross section \cite{RN460,RN494}:
\begin{eqnarray}
\sigma_\mathrm{cap}=\frac{10R_0^2\hbar\omega_0}{2E_{\mathrm{c.m.}}}\ln(1+\exp{\left[\frac{2\pi\left(E_{\mathrm{c.m.}}-E_0\right)}{\hbar\omega_0}\right]}).\label{eq:3}
\end{eqnarray}
In Eq.~(\ref{eq:3}), $E_0$ is the height of the total potential, and $\hbar\omega_0$ shows the inverted harmonic oscillator potential and is given as: 
\begin{eqnarray}
\hbar\omega_0=\frac{\hbar}{\sqrt\mu}\sqrt{\left|\frac{d^2V_\mathrm{Total}(r)}{dr^2}\right|_{R_l}},\label{eq:4}
\end{eqnarray}
where $V_\mathrm{Total}(r)$ is the total interaction potential and should be precisely calculated.
 
\subsection{Total interaction potential}

\begin{figure}
\includegraphics[width=58mm]{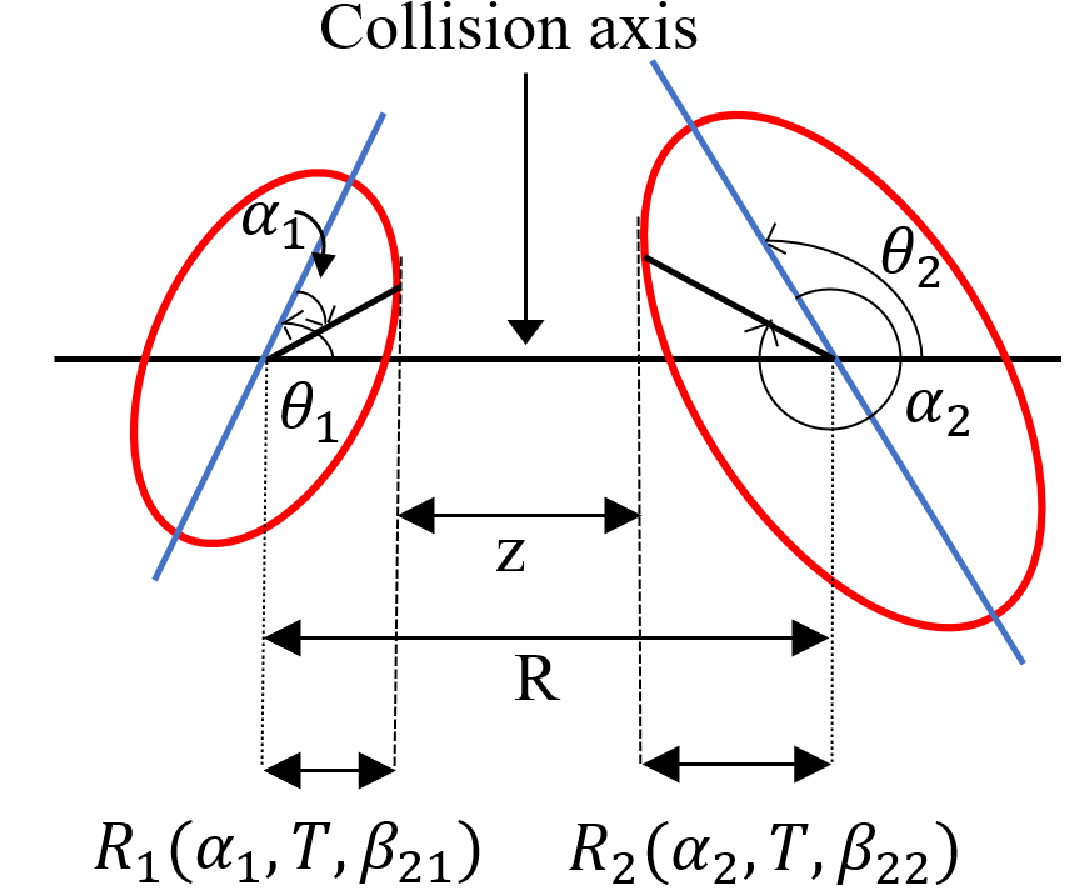}
\caption{\label{fig:fig1} Interaction angles of two nuclei during fusion.}
\end{figure}

The potential barrier characteristics, such as barrier height, position, curvature, and shape, are significant parameters that should be evaluated. Total potential is the function of two nuclei's radius, distance, colliding angle, and deformation parameters. Total potential is equal to the summation of the coulomb long-range repulsive potential, nuclear short-range attractive potential, and centrifugal potential \cite{RN496}:  
\begin{eqnarray}
	V_\mathrm{Total}=&&{V_C\left(R.Z_i,\beta_{\lambda i},\theta_i\right)+V}_N\left(R,A_i,\beta_{\lambda i},\theta_i\right)\nonumber\\+&&V_l(R,A_i,\beta_{\lambda i},\theta_i),\label{eq:5}
\end{eqnarray}
where $V_C\left(R.Z_i,\beta_{\lambda i},\theta_i\right)$ is the Coulomb potential, $V_N\left(R,A_i,\beta_{\lambda i},\theta_i\right)$ shows the nuclear potential, and $V_l(R,A_i,\beta_{\lambda i},\theta_i)$ denotes the centrifugal potential. Considering the multipole deformations, the Coulomb potential is shown as \cite{RN494,RN496}. 
\begin{eqnarray}
	&&V_C\left(R.Z_i,\beta_{\lambda i},\theta_i\right)=\frac{Z_1Z_2e^2}{r}+{3Z}_1Z_2e^2\nonumber\\&&\times\sum_{\lambda,i=1,2}{\frac{R_i^\lambda\left(\alpha_i\right)}{\left(2\lambda+1\right)R^{\lambda+1}}Y_\lambda^{(0)}(\theta_i)\left[\beta_{\lambda i}+\frac{4}{7}\beta_{\lambda i}^2Y_\lambda^{(0)}(\theta_i)\right]}.\nonumber\\\label{eq:6}
\end{eqnarray}
In Eq. ~(\ref{eq:6}), $Z_1 Z_2$ is the Coulomb factor and is related to the projectile and target, $r$ is the center of two nuclei distance, $R_i^\lambda\left(\alpha_i\right)$ is the deformed radii of the projectile and target, $\beta_{\lambda i}$ represents multipole deformations of the projectile and target, $\theta_i$ is the angle between the nuclear symmetry axis and the collision $Z$ axis, and $\alpha_i$ is the angle between the symmetry axis and the radius vector of the colliding nucleus \cite{RN497,RN529}. It should be noted that $\alpha_i$ is measured in the clockwise direction from the symmetry axis while $\theta_i$ is measured in the counterclockwise direction, as shown in Fig.~\ref{fig:fig1}.

The proximity potential is used to calculate the nuclear potential. This potential is calculated as follows \cite{RN499}:
\begin{eqnarray}
	V_N\left(R,A_i,\beta_{\lambda i},\theta_i\right)=4\pi\gamma\ bR\phi(\frac{z}{b}),\label{eq:7}
\end{eqnarray}
where the surface width $b$ is denoted \cite{RN499}. 
\begin{eqnarray}
	b=0.99.\label{eq:8}
\end{eqnarray}
In Eq.~(\ref{eq:7}) the central radius $R$ is obtained \cite{RN499}.
\begin{eqnarray}
	R=\frac{C_1{C}_2}{C_1+C_2},\label{eq:9}
\end{eqnarray}
where $C_i$ are Sussman central radii for the projectile and target and are calculated as follows:
\begin{eqnarray}
	C_i={R_i\left(1-\left(\frac{b}{R_i}\right)^2\right)}_{i=1,2},\label{eq:10}
\end{eqnarray}
In Eq.~(\ref{eq:7}), $\phi(\frac{z}{b})$ is the universal function where in $\xi=\frac{z}{b}=\frac{r-C_1-C_2}{b}$, $z$ is the minimum distance between projectile and target, and $\gamma$ shows the surface energy coefficient and is obtained from \cite{RN499}. 
\begin{eqnarray}
	\gamma=\gamma_0[1-k_s\frac{\left(N-Z\right)^2}{A^2}].\label{eq:11}
\end{eqnarray}
In Eq.~(\ref{eq:11}), $\gamma_0$ is the surface energy constant and $k_s$ is the surface asymmetric constant and are equal to $0.9180~\mathrm{MeV{fm}^{-2}}$ and $0.7546$, respectively \cite{RN491}. The universal function is obtained as \cite{RN499}.
\begin{eqnarray}
	\phi\left(\xi\right)=&&-1.7817+0.9270\xi+0.143\xi^2-0.09\xi^3,\nonumber\\ &&\mathrm{for}\ \xi\le0,\nonumber
\end{eqnarray}
\begin{eqnarray}
	\phi\left(\xi\right)=&&-1.7817+0.9270\xi+0.01696\xi^2-0.05148\xi^3,\nonumber\\ &&\mathrm{for}\ 0\le\xi\le1.9475,\label{eq:12}
\end{eqnarray}
\begin{eqnarray}
	\phi\left(\xi\right)=-4.41\mathrm{exp}\left(\frac{-\xi}{0.7176}\right), \mathrm{for}\ \xi\geq1.9475.\nonumber
\end{eqnarray}
Considering the nuclear deformations, the projectile and target radii are given as \cite{RN494}.
\begin{eqnarray}
	R_i\left(\alpha_i\right)=R_{0i}\left[1+\sum_{\lambda}{\beta_{\lambda i}Y_\lambda^{(0)}(\alpha_i)}\right].\label{eq:13}
\end{eqnarray}
Here, $R_{0i}$ denotes nuclear radii of the colliding participant and are given as \cite{RN499}.
\begin{eqnarray}
	R_{0i}=\left(1.28{A_i}^\frac{1}{3}-0.76+0.8{A_i}^{-\frac{1}{3}}\right)_{i=1,2}.\label{eq:14}
\end{eqnarray}
In Eq.~(\ref{eq:5}) centrifugal potential is given as.
\begin{eqnarray}
	V_l=\frac{\hbar^2l(l+1)}{2I_\mathrm{NS}}.\label{eq:15}
\end{eqnarray}
In Eq.~(\ref{eq:15}) $I_\mathrm{NS}=\ \mu\ r^2$ is the nonsticking moment of inertia, and $l$ is the angular momentum.\\
Considering the different collision angles in deformed nuclei, the capture cross section will be obtained by integrating the cross section over all collision angles \cite{RN564,RN496}:
\begin{eqnarray}
	\sigma_\mathrm{cap}=\int_{\theta_i=0}^{\frac{\pi}{2}}{\sigma(E_{\mathrm{c.m.}},\theta_i)sin\theta_1d\theta_1sin\theta_2d}\theta_2.\label{eq:16}
\end{eqnarray}

\subsection{The fusion cross section}

The fusion cross section is described as \cite{RN455}.
\begin{eqnarray}
	\sigma_\mathrm{fus}=\frac{\pi\hbar^2}{2\mu E_{\mathrm{c.m.}}}\sum_{l=0}^{l_{max}}{(2l+1)}T_l\left(E_{\mathrm{c.m.}},l\right)P_{CN}\left(E_{CN}^\ast,l\right),\nonumber\\\label{eq:17}
\end{eqnarray}
where $P_{CN}\left(E_{CN}^\ast,l\right)$ is the fusion probability, which is calculated as follows \cite{RN464}:   
\begin{eqnarray}
	P_{CN}\left(E_{CN}^\ast,l\right)=\frac{P_{CN}^0}{1+\mathrm{exp}\left[\frac{E_B^\ast-E_{CN}^\ast}{\Delta}\right]},\label{eq:18}
\end{eqnarray}
\begin{eqnarray}
	P_{CN}^0=\frac{1}{1+\mathrm{exp}\left[\frac{Z_1Z_2-\zeta}{\tau}\right]}.\label{eq:19}
\end{eqnarray}
In Eqs.~(\ref{eq:18}) and~(\ref{eq:19}), $E_{CN}^\ast$ is the CN excitation energy, $E_B^\ast$ is the energy of the CN when the energy of the center-of-mass is equal to the Bass barrier, $\Delta$ is an adjustable parameter and is often set around $4~\mathrm{MeV}$, and $\zeta~\approx1760$ and $\tau~\approx45$ are the fitted parameters \cite{RN464}:
\begin{eqnarray}
	E_{CN}^\ast=E_\mathrm{c.m.}+Q_\mathrm{val}\ \ .\label{eq:20}
\end{eqnarray}
In Eq.~(\ref{eq:20}), $E_{CN}^\ast$ is the CN excitation energy, $E_\mathrm{c.m.}$ denotes the system center-of-mass energy, and $Q_\mathrm{val}$ shows heat of reaction and is obtained as follows:
\begin{eqnarray}
	Q_\mathrm{val}=\mathrm{\Delta M}\left(A,Z\right)-\mathrm{\Delta M}\left(A_1,Z_1\right)-\mathrm{\Delta M}\left(A_2,Z_2\right),\nonumber\\\label{eq:21}
\end{eqnarray}
where $\mathrm{\Delta M}\left(A,Z\right)$, $\mathrm{\Delta M}\left(A_1,Z_1\right)$, and $\mathrm{\Delta M}\left(A_2,Z_2\right)$ are the mass excess values of the CN, projectile, and target, respectively. In this work, to calculate $Q_\mathrm{val}$, the mass excess was obtained from the Möller \textit{et al}. \cite{RN500} tables.\\
 
\subsection{The evaporation residue cross section}

Once the excited CN is formed, the equilibrium condition is achieved via the evaporation of one or more neutrons before decay. Therefore, one should consider the competition between neutron emission and decay to fission fragments in the CN. In hot fusion reactions, the formed CN has an excitation energy of $E_B^\ast>15~\mathrm{MeV}$, where the equilibrium condition is reached by the evaporation of two, three, or four neutrons involved \cite{RN455,RN531}. The evaporation residue cross section is given by.
\begin{eqnarray}
	\sigma_{ER}^{xn}=&&\frac{\pi\hbar^2}{2\mu E_{\mathrm{c.m.}}}\sum_{l=0}^{l_{max}}{\left(2l+1\right)T_l\left(E_{\mathrm{c.m.}},l\right)}\nonumber\\&&\times P_{CN}\left(E_{CN}^\ast,l\right)W_\mathrm{sur}^{xn}(E_{CN}^\ast,l),\label{eq:22}
\end{eqnarray}
where $W_\mathrm{sur}^{xn}(E_{CN}^\ast,l)$ denotes the survival probability of CN. Because of the high Coulomb barrier for the emission of charged particles from the excited heavy nucleus, the widths for the emission of a proton or an $\alpha$ particle are much smaller than the neutron emission width $\mathrm{\Gamma}_n$. Under these circumstances, ${\mathrm{\Gamma}_t\approx\mathrm{\Gamma}}_n+\mathrm{\Gamma}_f$ was set, and the survival probability $W_\mathrm{sur}^{xn}(E_{CN}^\ast,l)$ reflects the competition between neutron evaporation and fission of the excited CN. The survival probability is given by \cite{RN465}  
\begin{eqnarray}
	W_\mathrm{sur}^{xn}(E_{CN}^\ast,l)=P_{xn}(E_{CN}^\ast)\prod_{i=1}^{i_{max}=x}{(\frac{\mathrm{\Gamma}_n}{\mathrm{\Gamma}_n+\mathrm{\Gamma}_f})}_{i,E_{CN}^\ast}.\nonumber\\\label{eq:23}
\end{eqnarray}
In Eq.~(\ref{eq:23}) $\mathrm{\Gamma}_n$ and $\mathrm{\Gamma}_f$ are partial neutron emission width and partial fission width, respectively. At $E_{CN}^\ast>10~\mathrm{MeV}$, the $\mathrm{\Gamma}_n$ is much smaller than the $\mathrm{\Gamma}_f$. Therefore, the survival probability $W_\mathrm{sur}^{xn}(E_{CN}^\ast,l)$ is mainly determined by the realization probability $P_{xn}(E_{CN}^\ast)$ and the ratio $\frac{\mathrm{\Gamma}_n}{\mathrm{\Gamma}_f}$ \cite{RN458,RN501}:
\begin{eqnarray}
	&&\frac{\mathrm{\Gamma}_n}{\mathrm{\Gamma}_f}=\frac{4A^\frac{2}{3}a_f\left(E_{CN}^*-B_n\right)}{K_0a_n[2a_f^{\frac{1}{2}}\left[\left(E_{CN}^\ast-B_f\right)\right]^\frac{1}{2}-1)]}\nonumber\\&&\times \mathrm{exp}\left[2a_n^{1/2}\left(E_{CN}^*-B_n\right)^\frac{1}{2}-2a_f^{1/2}\left(E_{CN}^*-B_f\right)^\frac{1}{2}\right].\nonumber\\\label{eq:24}
\end{eqnarray}
In Eq.~(\ref{eq:24}), $B_n$ denotes the neutron binding energy, $B_f$ is the fission barrier, $K_0$ is set to a constant value of $10~\mathrm{MeV}$, and $a_n=A/10$ and $a_f=1.1a_n$ are the level density parameters of the fissioning nucleus at the ground state and saddle configurations, respectively \cite{RN501}. The fission barrier is given by \cite{RN456}
\begin{eqnarray}
	B_f\left(E_{CN}^\ast\right)=B_f^{LD}+S\mathrm{exp}{\left(\frac{-E_{CN}^\ast}{E_D}\right)}.\label{eq:25}
\end{eqnarray}
In Eq.~(\ref{eq:25}), $B_f^{LD}$ is the liquid drop (LD) fission barrier (macroscopic) and $S$ is the shell correction term . The liquid drop fission barrier is very low or equal to zero for heavy elements with $Z\geq109$ \cite{RN455,RN465}. The shell damping energy is given by \cite{RN530}. 
\begin{eqnarray}
	E_D=\frac{5.48A^\frac{1}{3}}{1+1.3A^{-\frac{1}{3}}}.\label{eq:26}
\end{eqnarray}
To calculate $P_{xn}(E_{CN}^\ast)$, in Eq.~(\ref{eq:23}) an equation which Jackson developed \cite{RN503} was employed. $P_{xn}(E_{CN}^\ast)$ represents the probability of emitting an exact number of neutrons, $x$, from the CN and is given by.
\begin{eqnarray}
	P_{xn}\left(E_{CN}^\ast\right)=I\left(\mathrm{\Delta}_x,2x-3\right)-\ I(\mathrm{\Delta}_{x+1},2x-1),\label{eq:27}
\end{eqnarray} 
where $I\left(z,n\right)$ is the Pearson’s incomplete $\mathrm{\gamma}$ function, and is obtained by
\begin{eqnarray}
	I\left(z,n\right)=\left(\frac{1}{n!}\right)\int_{0}^{z}m^ne^{-m}dm.\label{eq:28}
\end{eqnarray} 
In Eq.~ (\ref{eq:27}), $\mathrm{\Delta}_x=\frac{\left(E_{CN}^\ast-\sum_{1}^{x}{B_n(i)}\right)}{T}$, $B_n(i)$ is the binding energy of the $i\mathrm{th}$ evaporated neutron, and $T$ is the compound nucleus temperature and is described as:
\begin{eqnarray}
 E_{CN}^\ast=\frac{1}{a}{AT}^2-T,\label{eq:29}
\end{eqnarray}
where, $a$ represents the level density of CN and $A$ is the mass number of the CN \cite{RN501}.   
It should be noted that Eq.~(\ref{eq:27}) is valid for the calculation of the evaporation probability of two neutrons and more.
 
\begin{figure}
\includegraphics[width=85mm]{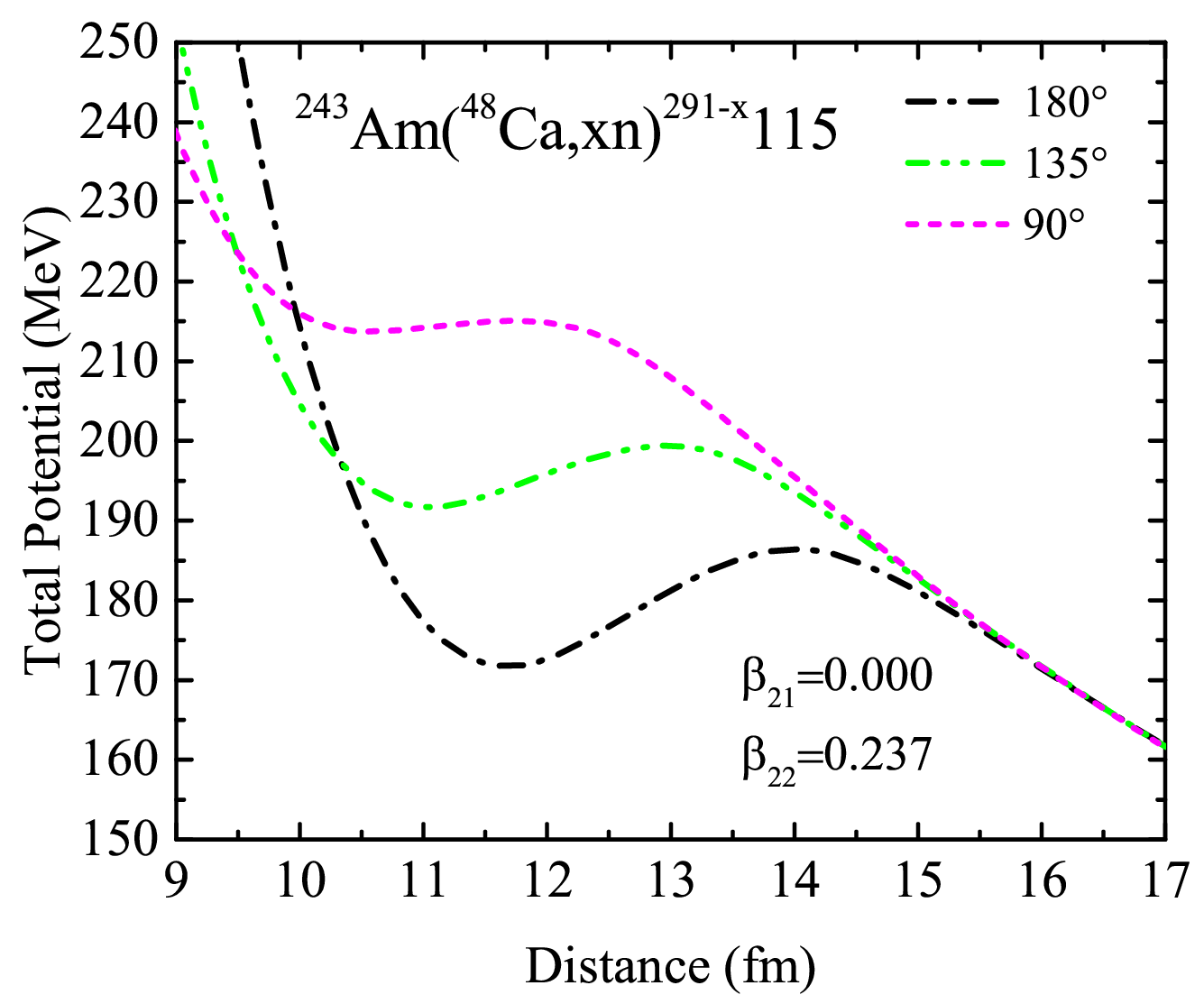}
\caption{\label{fig:fig2} Total potential vs distance for a combination of ${^{243}}\mathrm{Am}({^{48}}\mathrm{Ca},xn){^{291-x}}115$ is shown. It is noted that the spherical projectile ${^{48}}\mathrm{Ca}$ collides with oblate target ${^{243}}\mathrm{Am}$ under deformed nuclei, $\beta_{21}=0.237$. The dash-dot black lines show the potential at the target angle of 180\textdegree. The dash-dot-dot green lines show the potential at the target angle of 135\textdegree. The short-dash magenta lines show the potential at the target angle of 90\textdegree.}
\end{figure}

\begin{table}%The best place to locate the table environment is directly after its first reference in text
\caption{\label{tab:table1}
Combinations, calculated potential barriers, positions, and inverted harmonic oscillator potential.}
\setlength{\tabcolsep}{2pt}
\renewcommand{\arraystretch}{1.5}
\begin{ruledtabular}	
\begin{tabular}{lccc}
Combination&$V_B(\mathrm{MeV})$&$r_B(\mathrm{fm})$&$\hbar\omega_0(\mathrm{MeV})$\\ \hline
${^{48}}\mathrm{Ca}+{^{238}}\mathrm{U}$ & 193.65 & 12.94&3.353\\
${^{48}}\mathrm{Ca}+{^{237}}\mathrm{Np}$ & 196.17 & 12.90&3.328\\
${^{48}}\mathrm{Ca}+{^{240}}\mathrm{Pu}$ & 197.71 & 12.94&3.317\\
${^{48}}\mathrm{Ca}+{^{242}}\mathrm{Pu}$ & 1970.35 & 12.97&3.326\\
${^{48}}\mathrm{Ca}+{^{244}}\mathrm{Pu}$ & 197.00 & 13.00&3.334\\
${^{48}}\mathrm{Ca}+{^{243}}\mathrm{Am}$ & 199.38 & 12.97&3.308\\
${^{48}}\mathrm{Ca}+{^{245}}\mathrm{Cm}$ & 201.07 & 13.00&3.298\\
${^{48}}\mathrm{Ca}+{^{248}}\mathrm{Cm}$ & 200.53 & 13.04&3.303\\
${^{48}}\mathrm{Ca}+{^{249}}\mathrm{Bk}$ & 202.55 & 13.04&3.286\\
${^{48}}\mathrm{Ca}+{^{249}}\mathrm{Cf}$ & 204.74 & 13.03&3.271\\
\end{tabular}
\end{ruledtabular}
\end{table}

\begin{figure}
	\includegraphics[width=85mm]{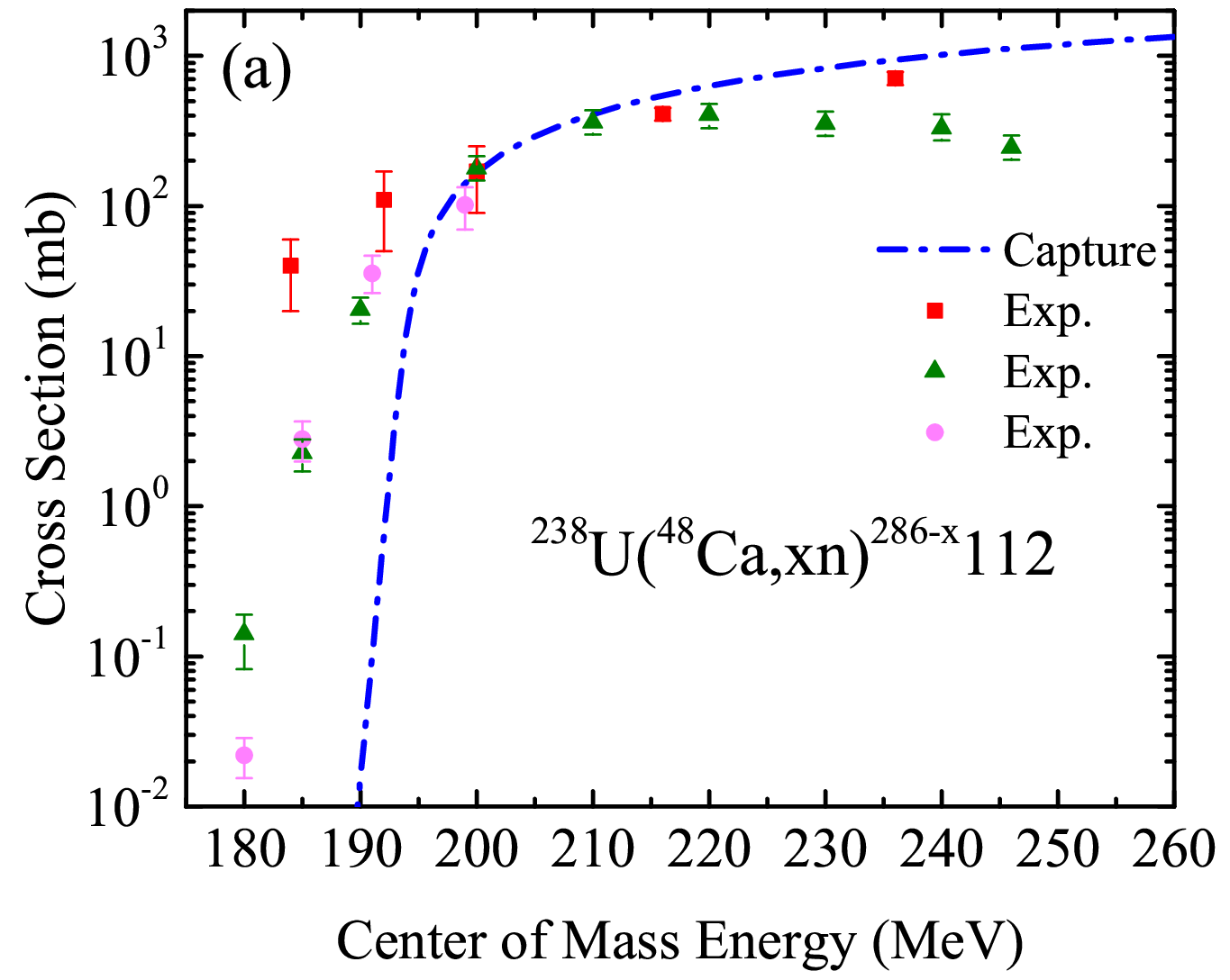}
	\includegraphics[width=85mm]{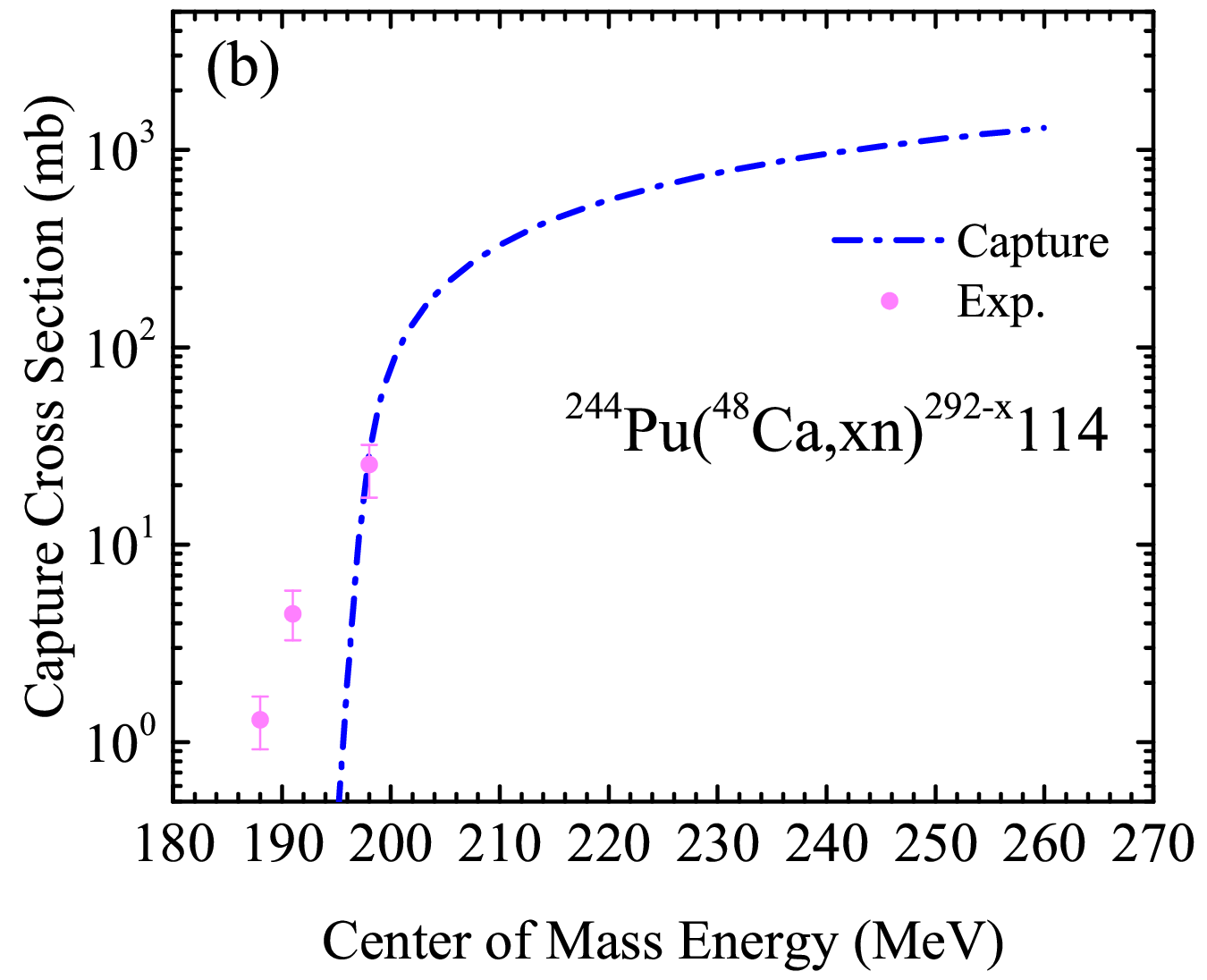}
	\caption{\label{fig:fig3} The capture cross section against the center of mass energy for combinations of (a)  ${^{48}}\mathrm{Ca}+{^{238}}\mathrm{U}$ and b) ${^{48}}\mathrm{Ca}+{^{244}}\mathrm{Pu}$. Experimental data obtained from \cite{RN559,RN560,RN561}.}
\end{figure}

\begin{figure}
	\includegraphics[width=85mm]{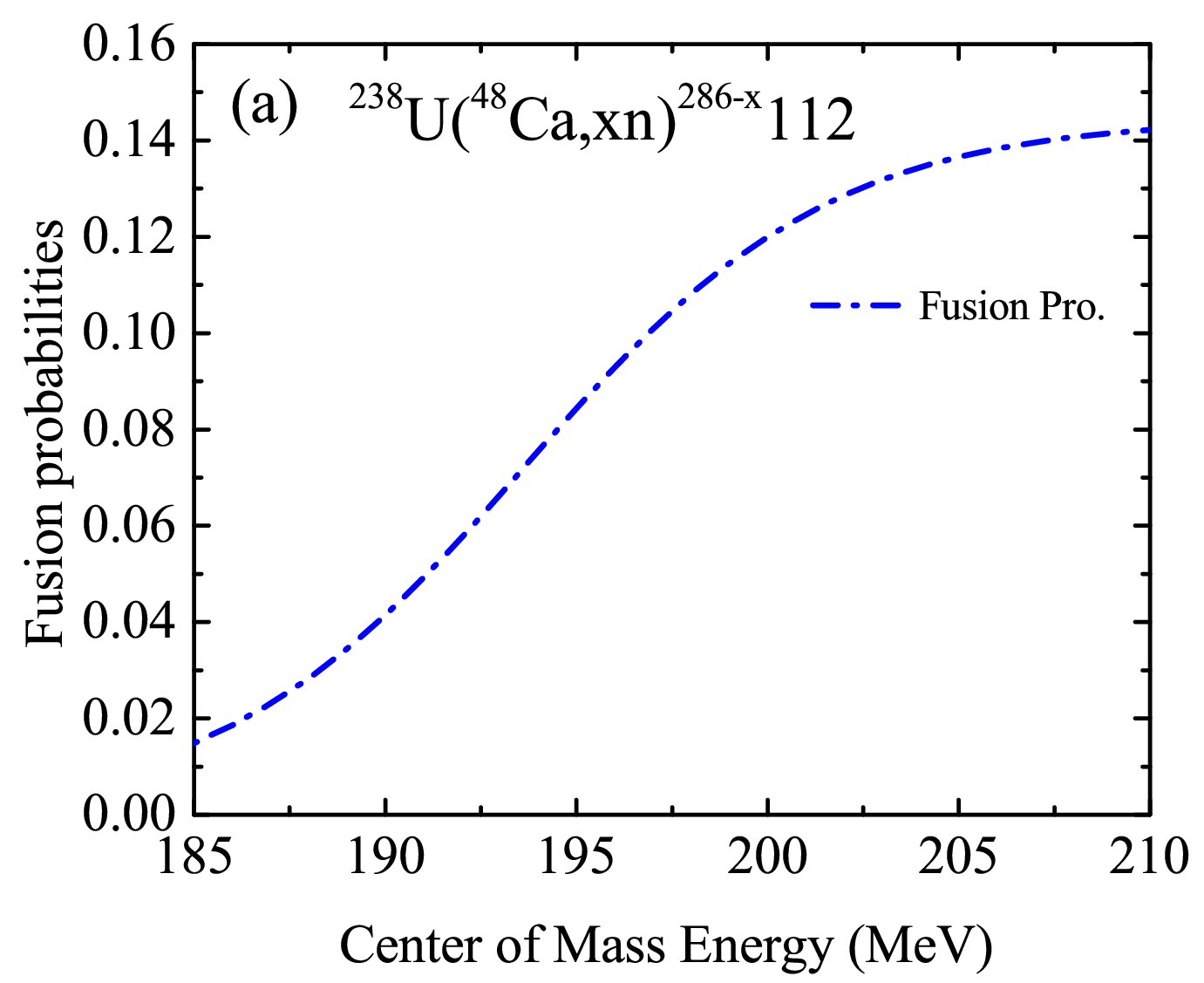}
	\includegraphics[width=85mm]{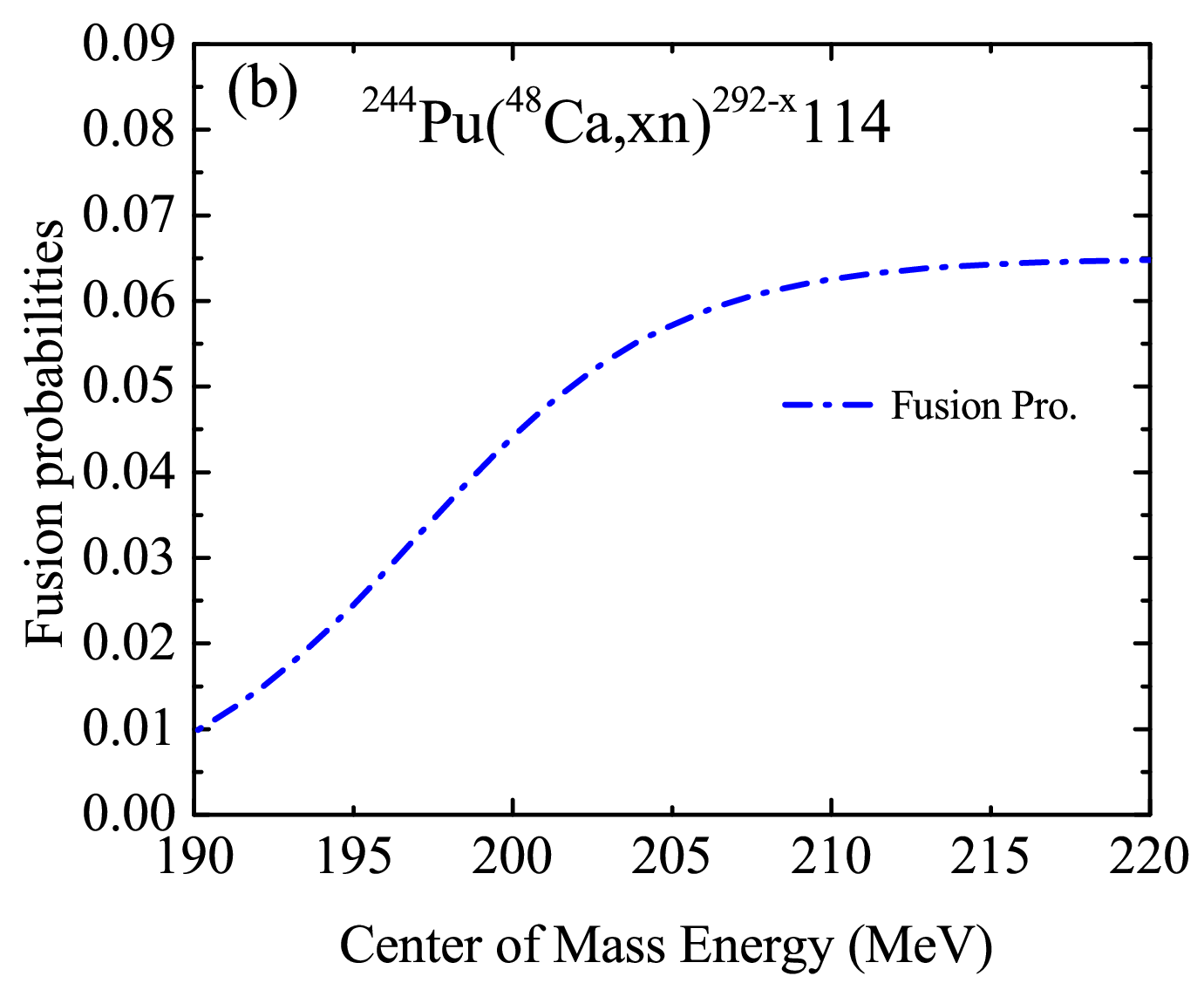}
	\caption{\label{fig:fig4} The fusion probability vs the center of mass energy for combinations of (a) ${^{48}}\mathrm{Ca}+{^{238}}\mathrm{U}$ and (b) ${^{48}}\mathrm{Ca}+{^{244}}\mathrm{Pu}$.}
\end{figure}

\begin{figure}
	\includegraphics[width=85mm]{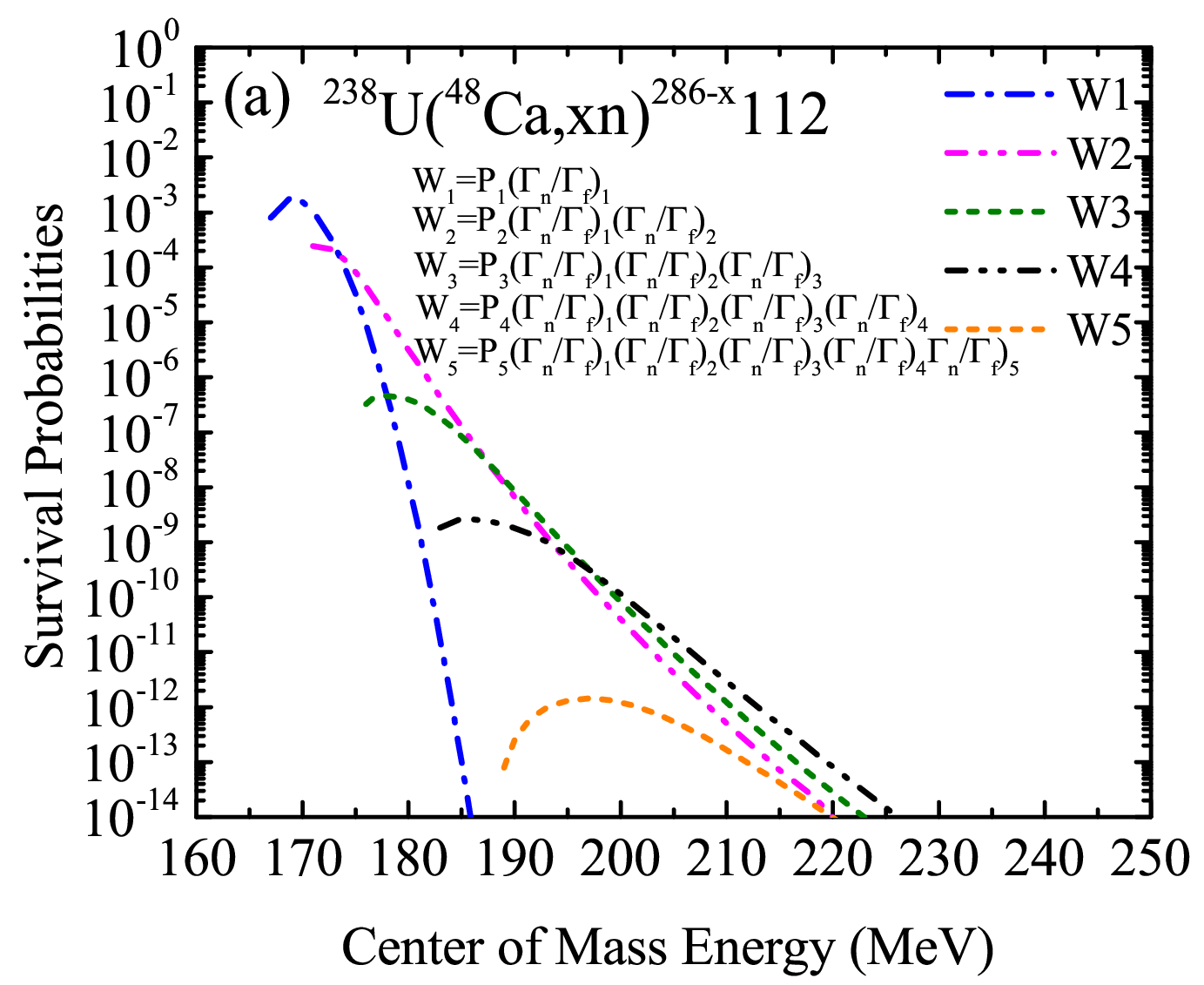}
	\includegraphics[width=85mm]{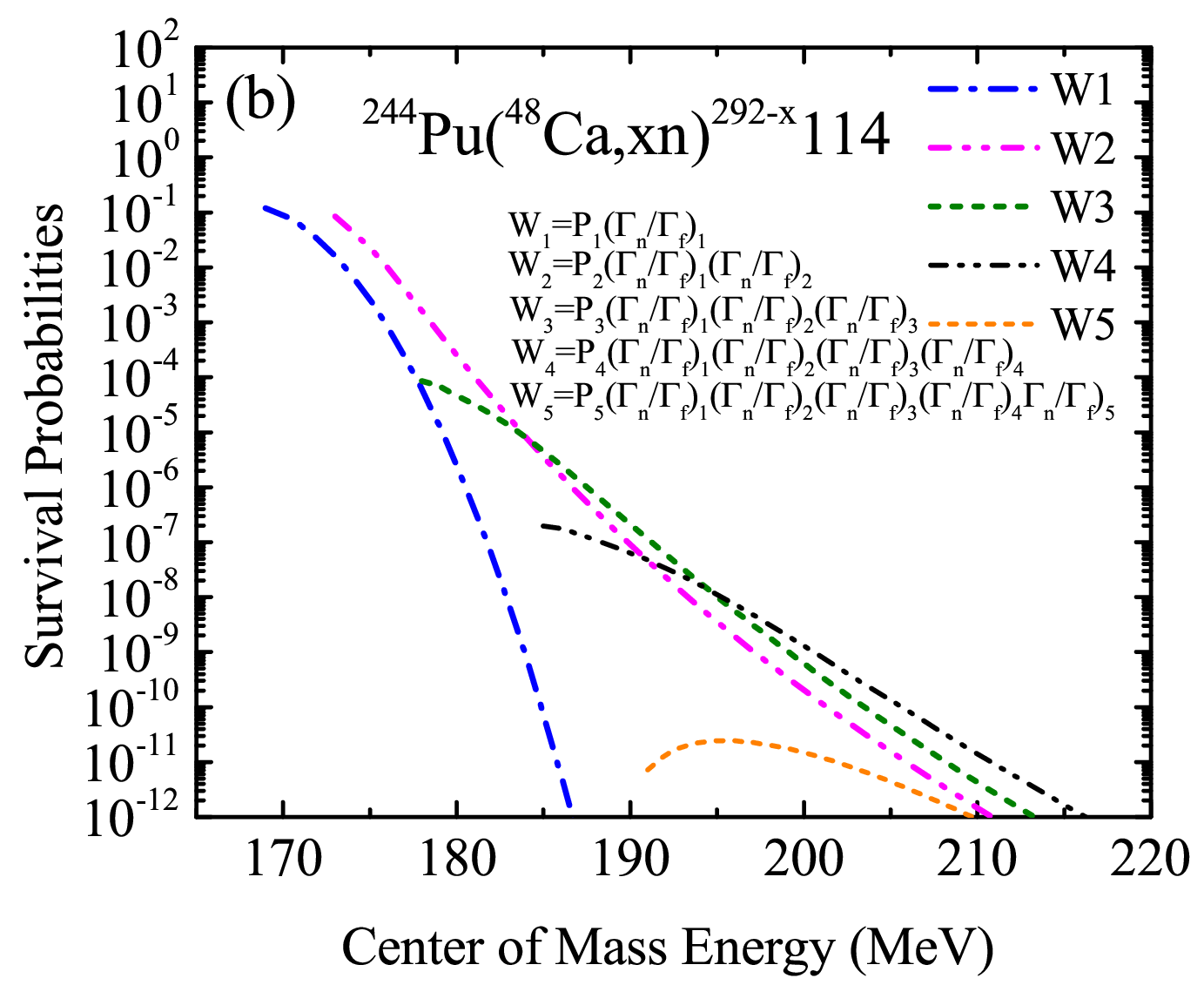}
	\caption{\label{fig:fig5} The survival probability vs the center of mass energy for combinations of (a) ${^{48}}\mathrm{Ca}+{^{238}}\mathrm{U}$ and (b) ${^{48}}\mathrm{Ca}+{^{244}}\mathrm{Pu}$.}
\end{figure}

\begin{table*}
	\caption{\label{tab:table2}Combination, the CN excitation energy, the center of mass energy, the ER cross section, the experimental CN excitation energy, and the ER cross section are shown. References give the experimental
		data \cite{RN472,RN473,RN474,RN475,RN476,RN477,RN478,RN479,RN480,RN481,RN482,RN483,RN484,RN485,RN486,RN487,RN565,RN566}.}
	\setlength{\tabcolsep}{2pt}
	\renewcommand{\arraystretch}{1.5}
	\begin{ruledtabular}	
		\begin{tabular}{cccccc}
			&\multicolumn{2}{c}{Theoretical calculation}&\multicolumn{2}{c}{Experimental data}&\\
			\cmidrule(rl){2-3} \cmidrule(rl){4-5}
			Combination & $E_{CN}^\ast[E_{\mathrm{c.m.}}](\mathrm{MeV})$ & $\sigma_{ER}(\mathrm{pb})$
			& $E_{CN}^\ast(\mathrm{MeV})$ & $\sigma_{ER}(\mathrm{pb})$ & Ref.\\ \hline
			${^{48}}\mathrm{Ca}+{^{238}}\mathrm{U}$ & 31.6[190] & $\sigma_{3n}=0.005$ & 29.3-33.5 & ${\sigma_{3n}=0.5}_{-0.41}^{+1.15}$ &\cite{RN472}\\
			& 34.6[193] & $\sigma_{3n}=0.51$ & 34.6 & ${\sigma_{3n}=0.7}_{-0.35}^{+0.58}$ & \cite{RN483}\\
			& 35.6[194] & $\sigma_{3n}=1.50$ & 32.9-37.2& ${\sigma_{3n}=2.5}_{-1.1}^{+1.8}$ & \cite{RN472}\\
			& 35.6[194] & $\sigma_{4n}=1.33$ & 32.9-37.2 & $\sigma_{4n}=0.8$ & \cite{RN472}\\
			& 36.6[195] & $\sigma_{3n}=2.54$ & 33.6-37.1& ${\sigma_{3n}=0.5}_{-0.3}^{+0.5}$ & \cite{RN566}\\
			& 39.6[198] & $\sigma_{4n}=2.81$ & 37.7-41.9 & ${\sigma_{4n}=0.6}_{-0.5}^{+1.6}$ & \cite{RN472}\\
			${^{48}}\mathrm{Ca}+{^{237}}\mathrm{Np}$ & 39.3[202] & $\sigma_{3n}=2.03$ & 36.9-41.2 & ${\sigma_{3n}=0.9}_{-0.6}^{+1.6}$ & \cite{RN473}\\
			${^{48}}\mathrm{Ca}+{^{240}}\mathrm{Pu}$ & 38.1[202] & $\sigma_{3n}=2.04$ & 36.5-41.1 &$\sigma_{3n}{=2.5}_{-1.4}^{+2.9}$ & \cite{RN484}\\
			& 43.1[207] & $\sigma_{4n}=0.65$ & 40.9-45.4 & ${\sigma_{4n}=2.6}_{-1.7}^{+3.3}$ & \cite{RN484}\\
			${^{48}}\mathrm{Ca}+{^{242}}\mathrm{Pu}$ & 32.8[195] & $\sigma_{2n}=0.02$ & 30.4-34.7 & ${\sigma_{2n}=0.5}_{-0.4}^{+1.4}$ & \cite{RN472}\\
			& 38.8[201] & $\sigma_{3n}=2.18$ & 37.1-40.7 & $\sigma_{3n}{=10.4}_{-2.1}^{+3.5}$ & \cite{RN566}\\
			& 38.8[201] & $\sigma_{4n}=2.60$ & 37.1-40.7 & $\sigma_{4n}{=1.8}_{-0.6}^{+1.0}$ & \cite{RN566}\\
			& 39.8[202] & $\sigma_{3n}=1.77$ & 38-42.4 & $\sigma_{3n}{=1.4}_{-1.2}^{+3.2}$ & \cite{RN481}\\
			& 40.8[203] & $\sigma_{4n}=2.02$ & 38-42.4 & $\sigma_{4n}{=1.4}_{-1.2}^{+3.2}$ & \cite{RN481}\\
			& 39.8[202] & $\sigma_{3n}=1.77$ & 38-42.4 & $\sigma_{3n}{=3.6}_{-1.7}^{+3.4}$ & \cite{RN472}\\
			& 40.8[203] & $\sigma_{4n}=2.02$ & 38-42.4 & ${\sigma_{4n}=4.5}_{-1.9}^{+3.6}$ & \cite{RN472}\\
			& 42.8[205] & $\sigma_{3n}=0.74$ & 41.3-44.8 & $\sigma_{3n}{=1.2}_{-0.7}^{+1.2}$ & \cite{RN566}\\
			& 42.8[205] & $\sigma_{4n}=1.29$ & 41.3-44.8 & $\sigma_{4n}{=4.8}_{-1.6}^{+2.1}$ & \cite{RN566}\\
			& 49.8[212] &	$\sigma_{4n}=0.14$	& 48-52 &	${\sigma_{4n}=0.6}_{-0.5}^{+0.9}$ &\cite{RN480}\\
			& 50.8[213] &	$\sigma_{5n}=0.006$ & 48-52 &	${\sigma_{5n}=0.6}_{-0.5}^{+0.9}$ &\cite{RN480}\\
			${^{48}}\mathrm{Ca}+{^{244}}\mathrm{Pu}$ & 37.4[198] & $\sigma_{3n}=1.93$ & 36.1-39.5 & $\sigma_{3n}{=8.0}_{-4.5}^{+7.4}$ &\cite{RN482}\\
			& 41.4[202] &	$\sigma_{3n}=1.41$ & 39.8-43.9 & $\sigma_{3n}{=3.5}_{-2.0}^{+3.3}$	&\cite{RN478}\\
			& 42.4[203] & $\sigma_{3n}=1.06$ & 39-43 & $\sigma_{3n}{=1.7}_{-1.1}^{+2.5}$ & \cite{RN486}\\
			& 37.4[198] &	$\sigma_{4n}=3.32$ & 36.1-40.1 & ${\sigma_{4n}=2.8}_{-2.1}^{+4.2}$	&\cite{RN478}\\
			& 40.4[201] &	$\sigma_{4n}=4.10$ & 39.8-43.9 & ${\sigma_{4n}=11}_{-7}^{+15}$	&\cite{RN478}\\
			& 41.4[202] &	$\sigma_{4n}=3.46$ & 39.8-43.9 & ${\sigma_{4n}=9.8}_{-3.1}^{+3.9}$	&\cite{RN482}\\
			& 42.4[203] &	$\sigma_{4n}=2.73$ & 39-43 & ${\sigma_{4n}=5.3}_{-2.1}^{+3.6}$	&\cite{RN486}\\
			& 52.4[213] &	$\sigma_{5n}=0.008$ & 50.4-54.7 &	${\sigma_{5n}=1.1}_{-0.9}^{+2.6}$	&\cite{RN486}\\
			${^{48}}\mathrm{Ca}+{^{243}}\mathrm{Am}$ &34.5[200] &	$\sigma_{2n}=0.92$ & 33.1-35.2 &$\sigma_{2n}={1.1}_{-0.9}^{+2.5}$ &\cite{RN565}\\
			&34.5[201] &	$\sigma_{3n}=2.76$ & 33.1-35.2 &$\sigma_{3n}={17.1}_{-4.7}^{+6.3}$ &\cite{RN565}\\
			&36.5[202] &	$\sigma_{2n}=1.23$ & 35.5-37.8 &$\sigma_{2n}={1.6}_{-0.7}^{+1.2}$ &\cite{RN565}\\
			&36.5[202] &	$\sigma_{3n}=3.89$ & 34-38.3 &	${\sigma_{3n}=8.5}_{-3.7}^{+6.4}$ &\cite{RN487}\\
			& 39.5[205] &	$\sigma_{3n}=2.06$ & 38-43.2 & ${\sigma_{3n}=3.7}_{-1.0}^{+1.3}$ &\cite{RN474}\\
			& 40.5[206] &	$\sigma_{3n}=1.51$ & 38-43.2 &	${\sigma_{3n}=2.7}_{-1.6}^{+4.8}$ &\cite{RN474}\\
			& 44.5[210] &	$\sigma_{4n}=0.53$ & 42.4-46.5 & ${\sigma_{4n}=0.9}_{-0.8}^{+3.2}$ &\cite{RN474}\\
			${^{48}}\mathrm{Ca}+{^{245}}\mathrm{Cm}$ & 32.7[201] &	$\sigma_{2n}=0.76$ & 30.9-35 &	${\sigma_{2n}=0.9}_{-0.6}^{+1.1}$ &\cite{RN486}\\
			& 37.7[206] &	$\sigma_{2n}=0.96$ & 35.9-39.9 & ${\sigma_{2n}=0.7}_{-0.6}^{+2.0}$	&\cite{RN475}\\
			& 32.7[201] &	$\sigma_{3n}=11.63$ & 30.9-35 & ${\sigma_{3n}=1.3}_{-0.7}^{+1.2}$	&\cite{RN486}\\
			
		\end{tabular}
	\end{ruledtabular}
\end{table*}

\begin{table*}
	Continued table~\ref{tab:table2}\\\hspace{-2.3cm}
	\setlength{\tabcolsep}{2pt}
	\renewcommand{\arraystretch}{1.5}
	\begin{ruledtabular}
		\begin{tabular}{cccccc}
			&\multicolumn{2}{c}{Theoretical calculation}&\multicolumn{2}{c}{Experimental-data}&\\
			\cmidrule(rl){2-3} \cmidrule(rl){4-5}
			Combination & $E_{CN}^\ast[E_{\mathrm{c.m.}}](\mathrm{MeV})$ & $\sigma_{ER}(\mathrm{pb})$
			& $E_{CN}^\ast(\mathrm{MeV})$ & $\sigma_{ER}(\mathrm{pb})$ & Ref.\\ \hline
			& 41.7[210] &	$\sigma_{3n}=4.06$ & 40.7-44.8 & ${\sigma_{3n}=1.9}_{-1.0}^{+2.1}$	&\cite{RN475}\\
			& 42.7[211] &	$\sigma_{3n}=2.75$ & 40.7-44.8 & ${\sigma_{3n}=3.7}_{-1.8}^{+3.6}$	&\cite{RN475}\\
			& 42.7[211] &	$\sigma_{4n}=0.64$ & 40.7-44.8 & ${\sigma_{4n}=1.0}_{-0.0}^{+0.0}$	&\cite{RN475}\\
			${^{48}}\mathrm{Ca}+{^{248}}\mathrm{Cm}$ & 30.4[197] & $\sigma_{3n}=0.014$ & 30.5 &	$\sigma_{3n}=0.9$&\cite{RN472}\\
			& 33.4[200] &	$\sigma_{3n}=0.9$ & 33 & ${\sigma_{3n}=0.5}_{-0.26}^{+0.5}$ &\cite{RN472}\\
			& 38.4[205] &	$\sigma_{3n}=2.38$ & 36.8-41.1 & ${\sigma_{3n}=1.1}_{-0.7}^{+1.7}$	&\cite{RN485}\\
			& 40.4[207] & $\sigma_{3n}=1.29$ & 40.9 &	${\sigma_{3n}=0.9}_{-0.7}^{+2.1}$ &\cite{RN485}\\
			& 33.4[200] &	$\sigma_{4n}=0.7$ & 33 & $\sigma_{4n}=0.3$	&\cite{RN472}\\
			& 38.4[205] &	$\sigma_{4n}=4.36$ & 36.8-41.1 & ${\sigma_{4n}=3.3}_{-1.4}^{+2.5}$	&\cite{RN485}\\
			& 40.4[207] & $\sigma_{4n}=2.84$ & 40.9 & ${\sigma_{4n}=3.4}_{-1.6}^{+2.7}$	&\cite{RN485}\\
			${^{48}}\mathrm{Ca}+{^{249}}\mathrm{Bk}$ & 32.8[203] &	$\sigma_{3n}=3.70$ & 30.4-34.7 & $\sigma_{3n}{=0.7}_{-0.57}^{+1.7}$	&\cite{RN477}\\
			& 34.8[205] & $\sigma_{3n}=5.45$ & 33.2-37.4 & $\sigma_{3n}{=0.5}_{-0.4}^{+1.1}$	&\cite{RN476}\\
			& 34.8[205] &	$\sigma_{3n}=5.45$ & 35 & $\sigma_{3n}={3.6}_{-2.5}^{+6.1}$ &\cite{RN479}\\
			& 34.8[205] &	$\sigma_{3n}=5.45$ & 32.8-37.5 & ${\sigma_{3n}=1.1}_{-0.6}^{+1.2}$	&\cite{RN477}\\
			& 38.8[209] &	$\sigma_{3n}=1.94$ &39 & $\sigma_{3n}=0.7$	&\cite{RN476}\\
			& 38.8[209] &	$\sigma_{3n}=1.94$ & 39 & $\sigma_{3n}=0.32$ &\cite{RN477}\\
			& 34.8[205] & $\sigma_{4n}=3.43$ & 35 & $\sigma_{4n}=0.8$ &\cite{RN476}\\
			& 34.8[205] &	$\sigma_{4n}=3.43$ & 35 & $\sigma_{4n}=0.59$ &\cite{RN477}\\
			& 38.8[209] &	$\sigma_{4n}=2.29$ & 37.2-41.4 & $\sigma_{4n}={1.3}_{-0.6}^{+1.5}$	&\cite{RN476}\\
			& 38.8[209] &	$\sigma_{4n}=2.29$ & 39 & ${\sigma_{4n}=2.0}_{-1.1}^{+2.2}$ &\cite{RN479}\\
			& 38.8[209] &	$\sigma_{4n}=2.29$ & 37-41.9 &	$\sigma_{4n}={1.5}_{-0.5}^{+1.1}$ &\cite{RN477}\\
			& 42.8[213] &	$\sigma_{4n}=0.71$ & 40.3-44.8 & $\sigma_{4n}={2.4}_{-1.4}^{+3.3}$	&\cite{RN477}\\
			& 45.8[216] &	$\sigma_{4n}=0.25$ & 43.8-48.3 & ${\sigma_{4n}=2.0}_{-1.1}^{+1.8}$	&\cite{RN477}\\
			${^{48}}\mathrm{Ca}+{^{249}}\mathrm{Cf}$ & 29.6[204] & $\sigma_{3n}=3.77$ & 26.6-31.7 & ${\sigma_{3n}=0.3}_{-0.27}^{+1.0}$	&\cite{RN475}\\
			& 34.6[209] &	$\sigma_{3n}=15.93$ & 32.1-36.6 & ${\sigma_{3n}=0.5}_{-0.3}^{+1.6}$ &\cite{RN475}\\
		\end{tabular}
	\end{ruledtabular}
\end{table*}

\begin{figure*}
\includegraphics[width=59mm]{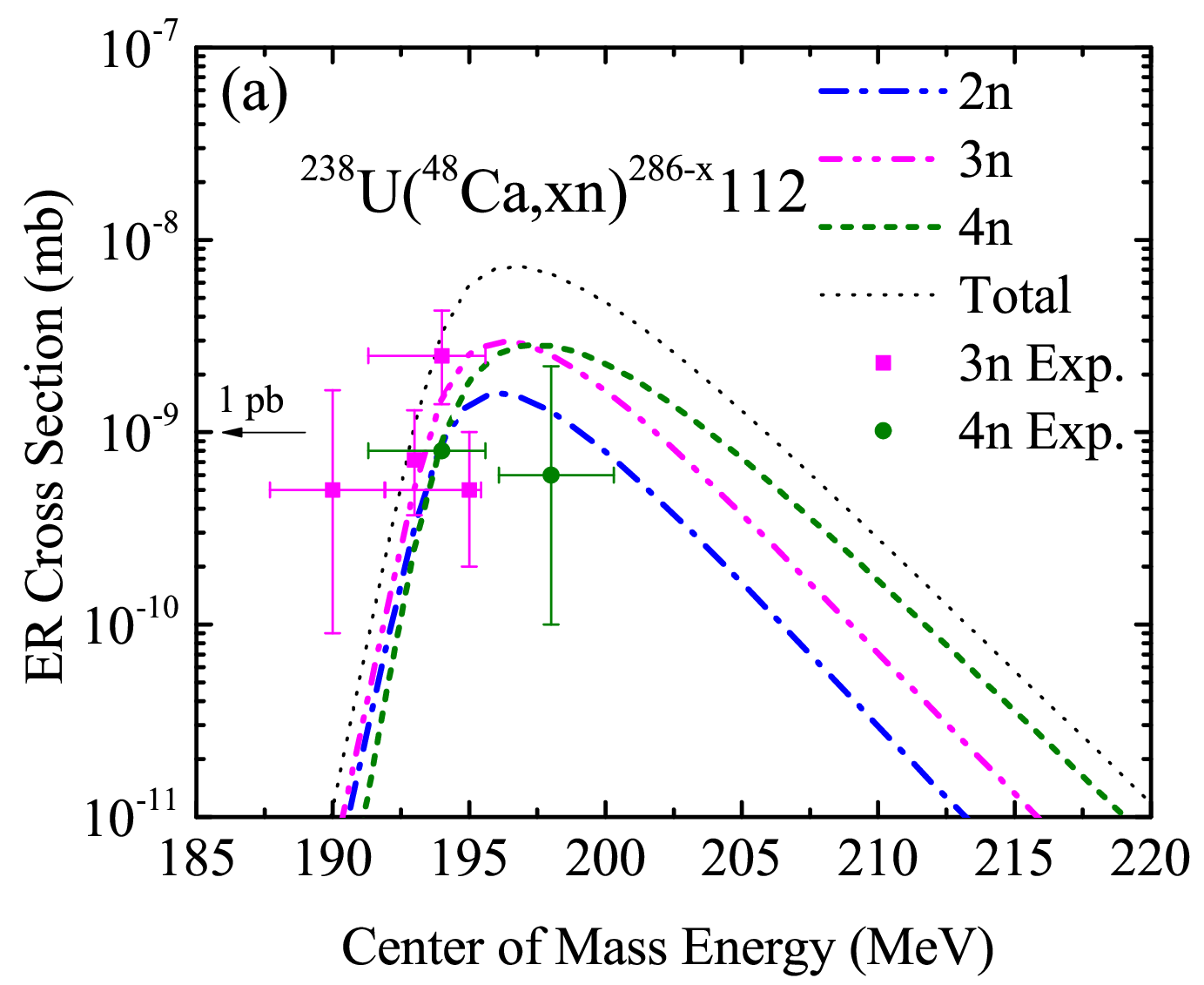}
\includegraphics[width=59mm]{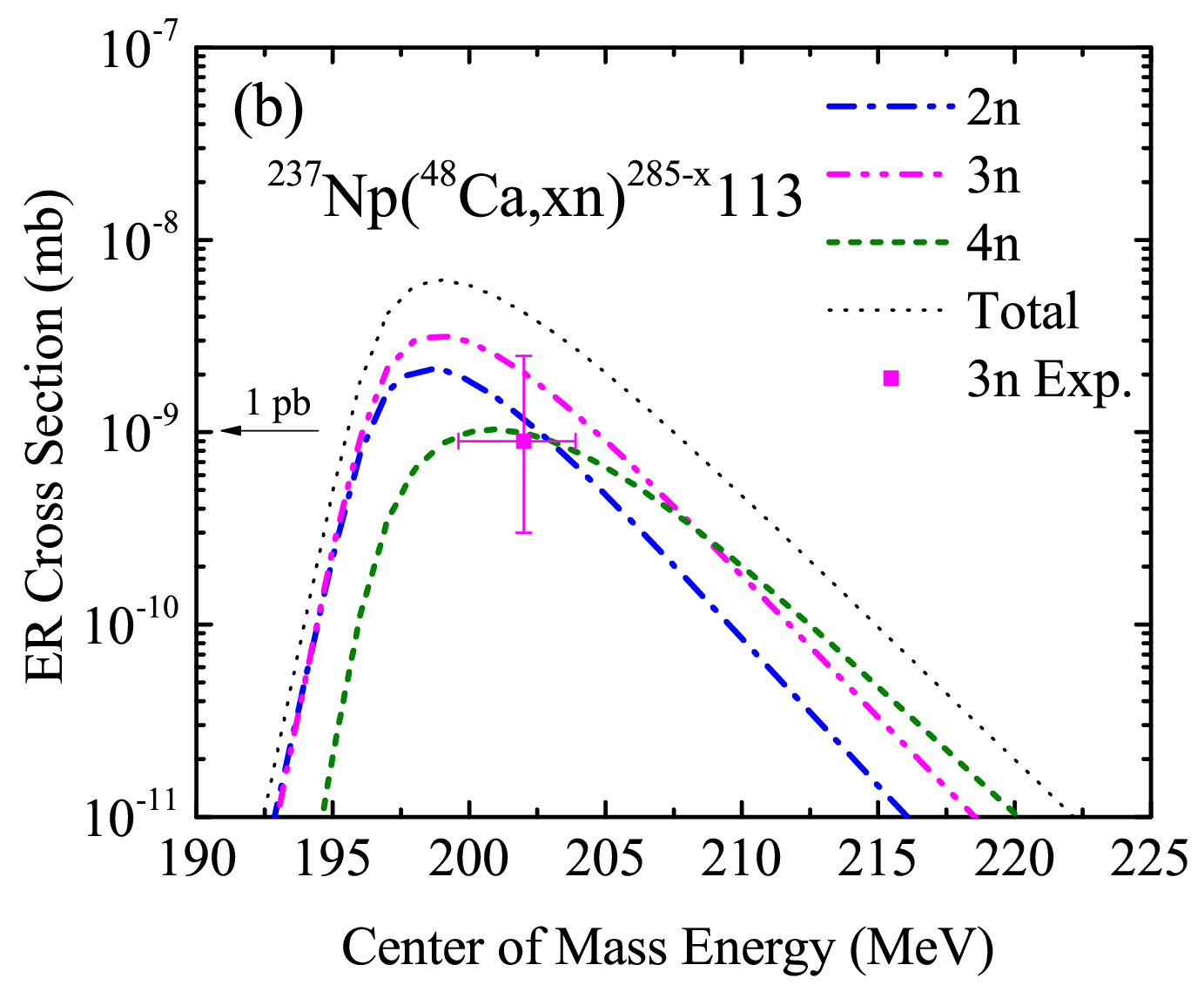}
\includegraphics[width=59mm]{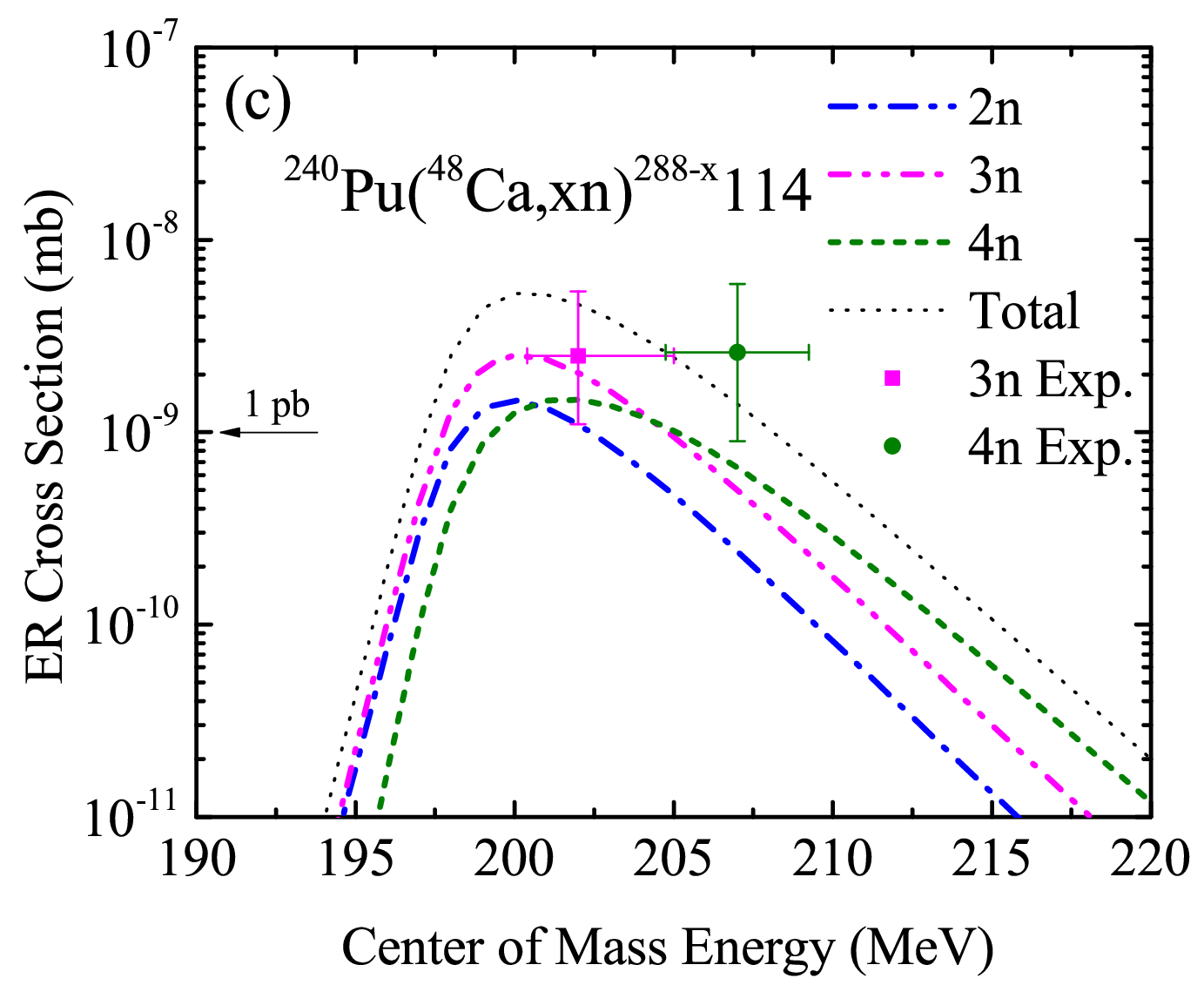}
\includegraphics[width=59mm]{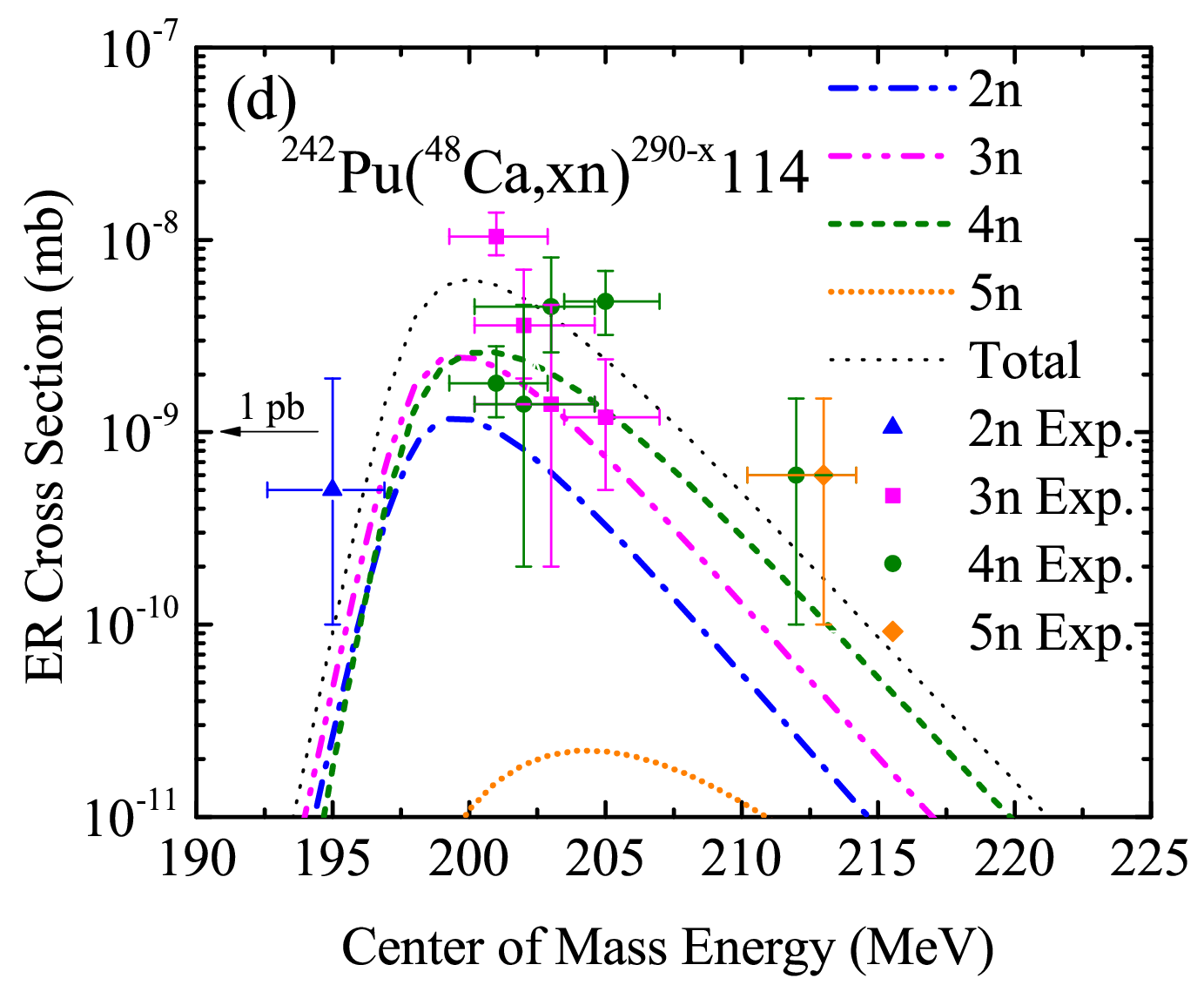}
\includegraphics[width=59mm]{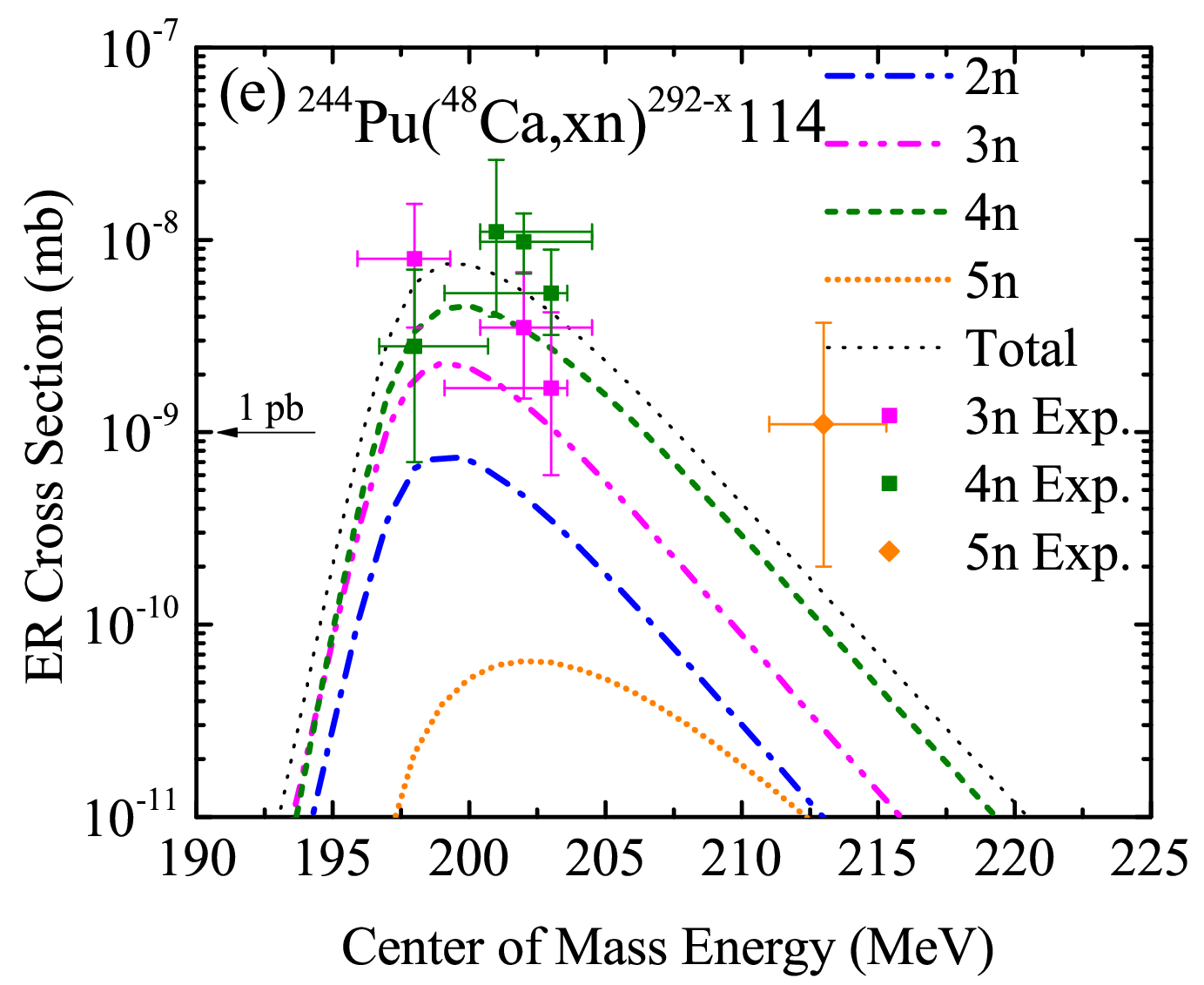}
\includegraphics[width=59mm]{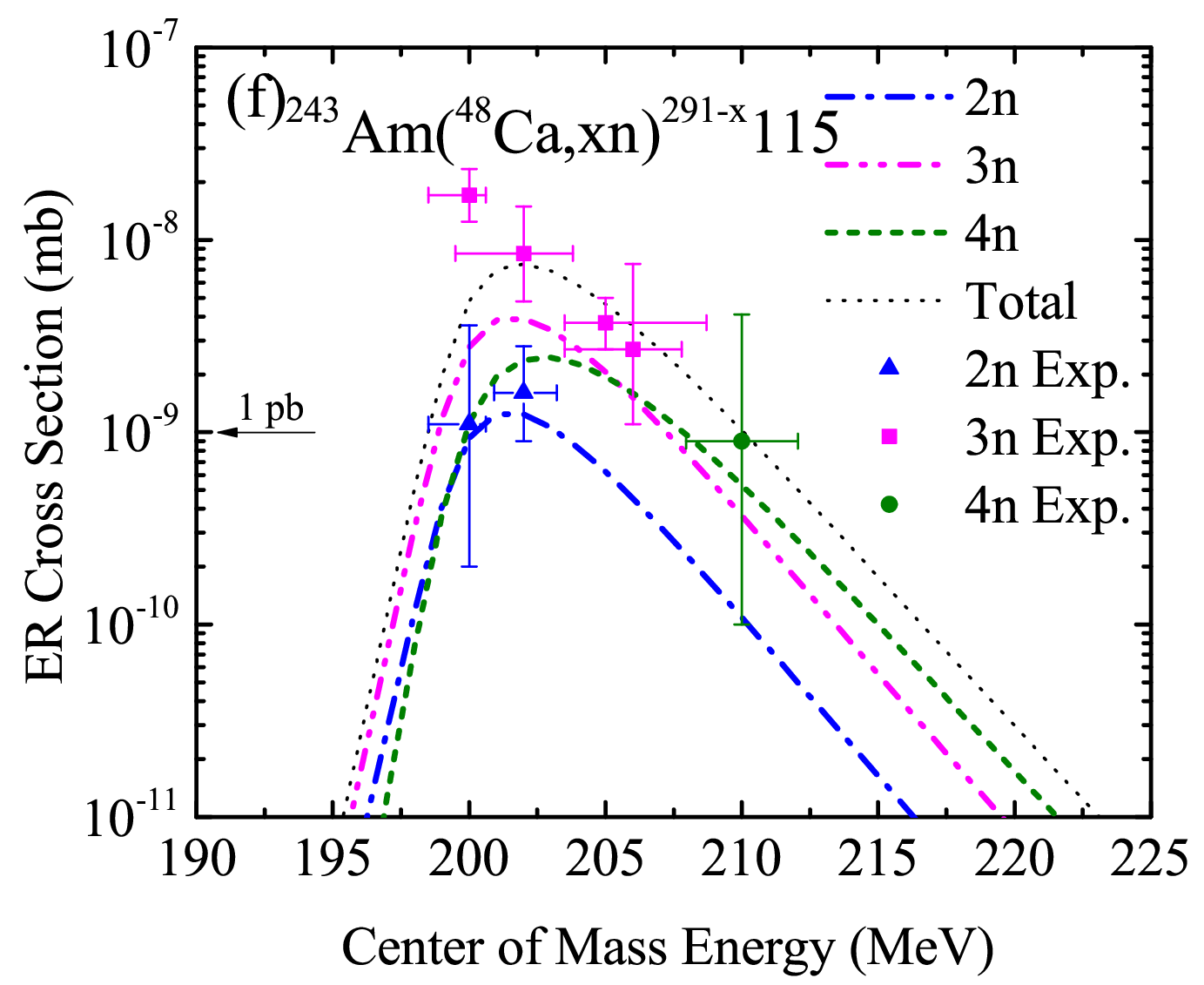}
\includegraphics[width=59mm]{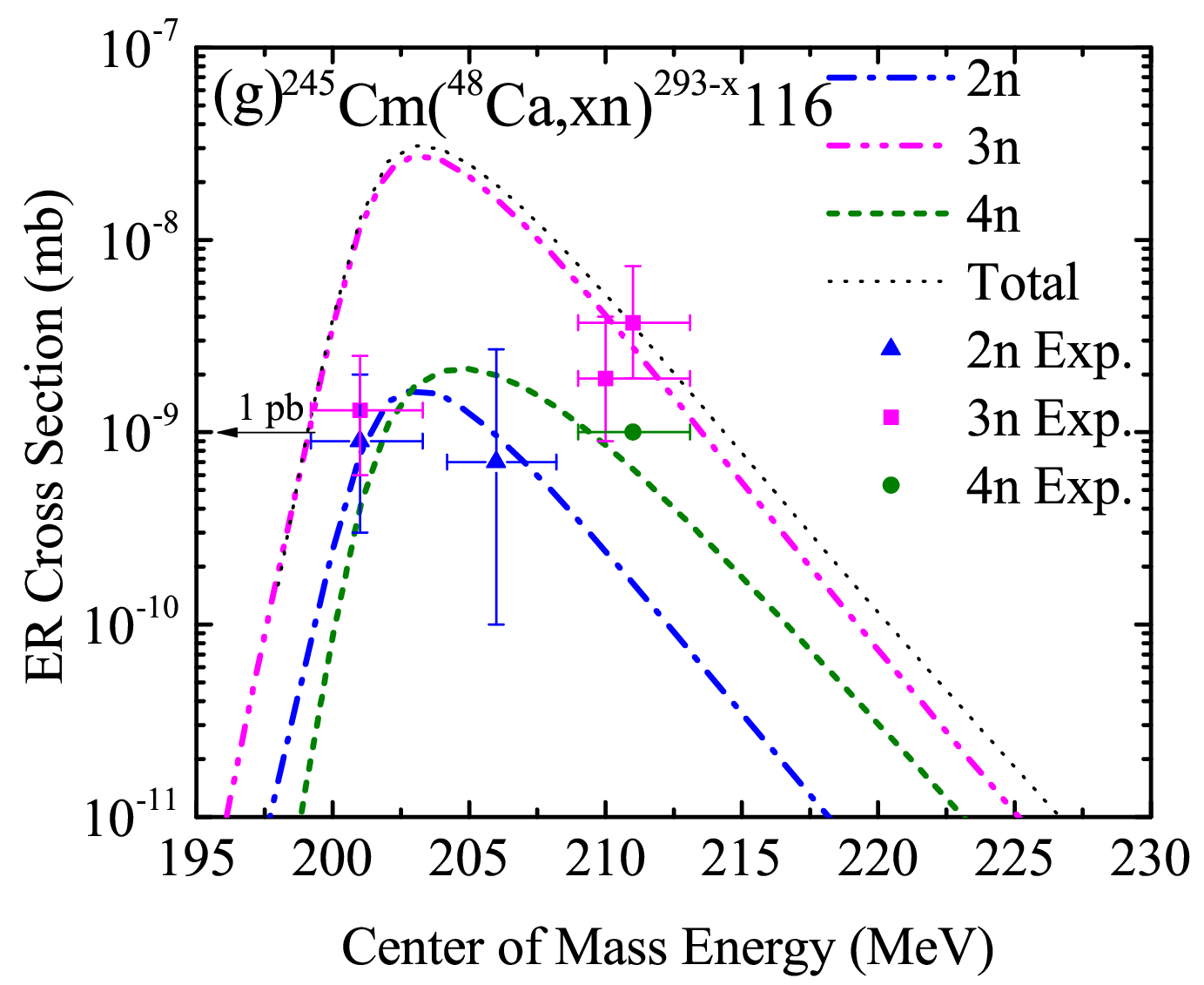}
\includegraphics[width=59mm]{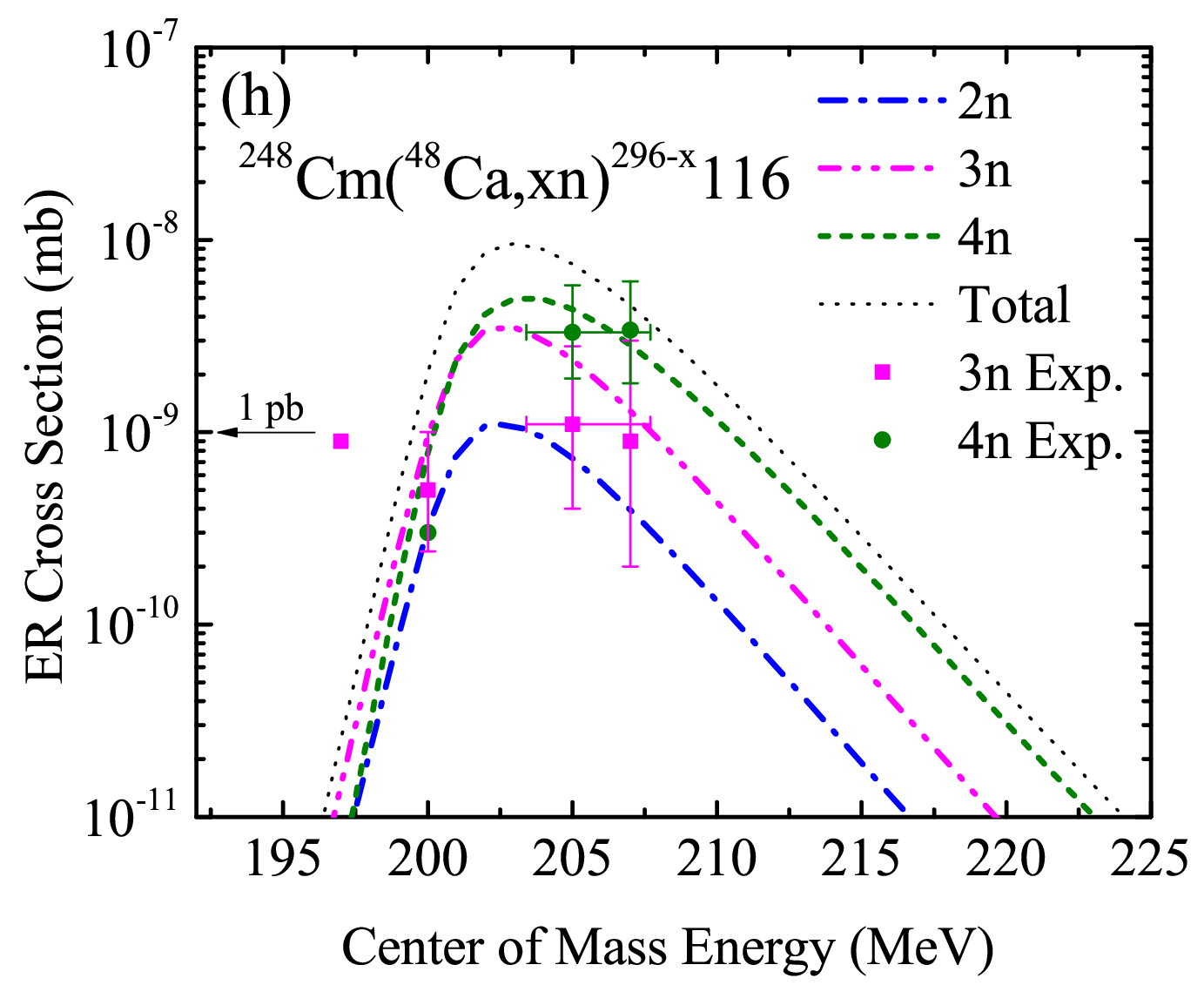}
\includegraphics[width=59mm]{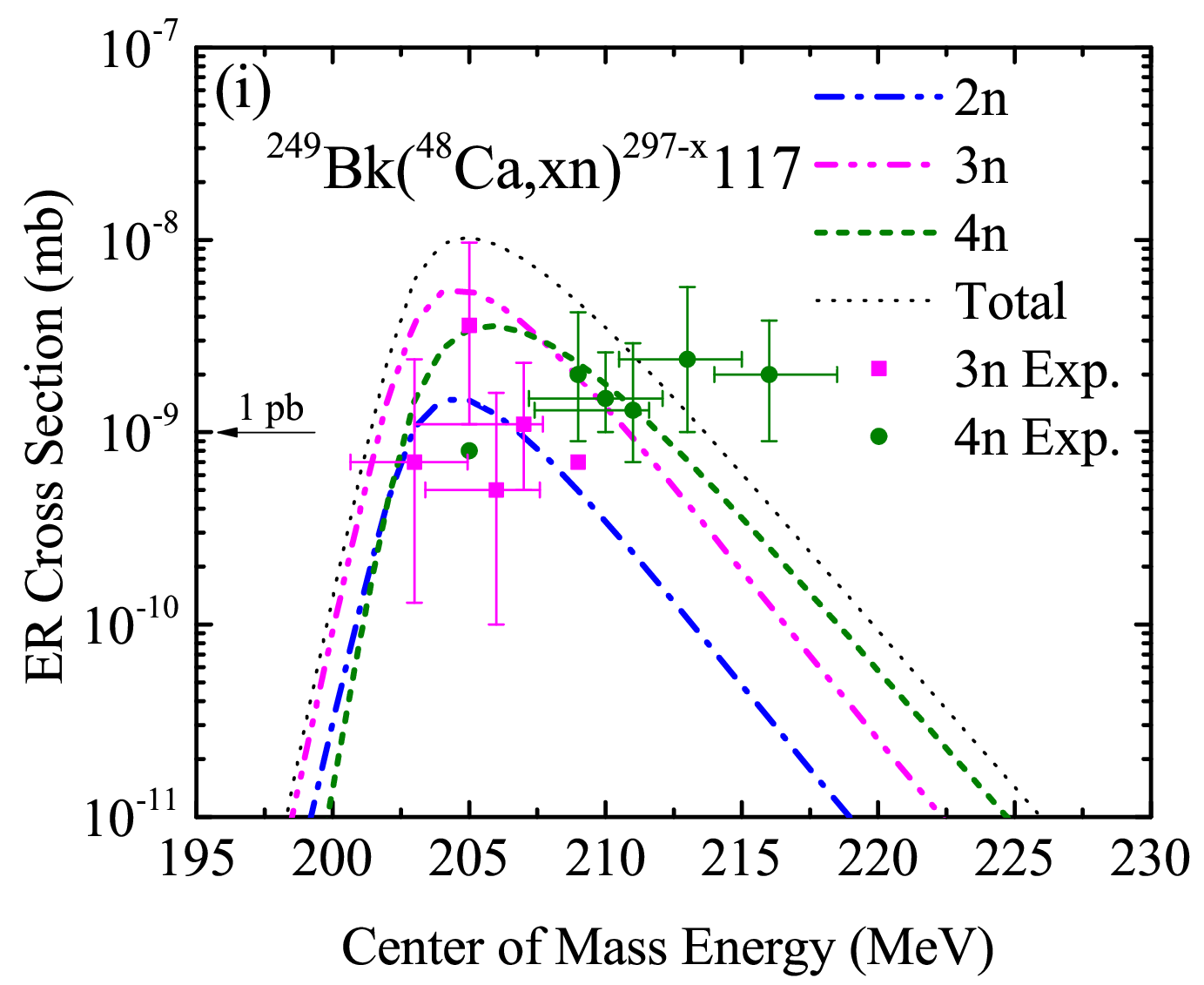}
\includegraphics[width=59mm]{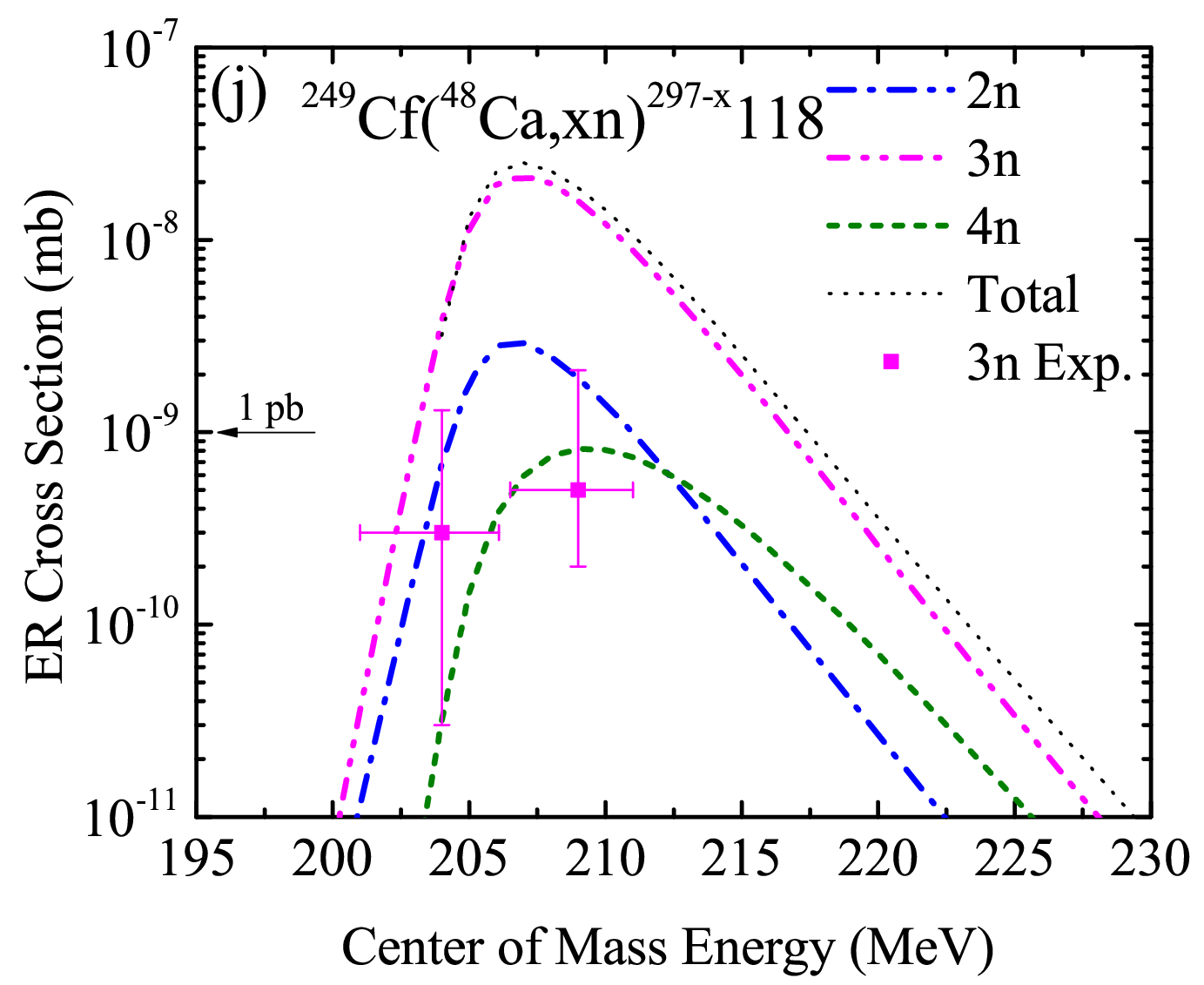}
\caption{\label{fig:fig6} The ER cross section vs the center of mass energy for combinations of  a) ${^{238}}\mathrm{U}({^{48}}\mathrm{Ca},xn){^{286-x}}112$, b) ${^{237}}\mathrm{Np}({^{48}}\mathrm{Ca},xn){^{285-x}}113$, (c) ${^{240}}\mathrm{Pu}({^{48}}\mathrm{Ca},xn){^{288-x}}114$, (d) ${^{242}}\mathrm{Pu}({^{48}}\mathrm{Ca},xn){^{290-x}}114$, (e) ${^{244}}\mathrm{Pu}({^{48}}\mathrm{Ca},xn){^{292-x}}114$, (f) ${^{243}}\mathrm{Am}({^{48}}\mathrm{Ca},xn){^{291-x}}115$, (g) ${^{245}}\mathrm{Cm}({^{48}}\mathrm{Ca},xn){^{293-x}}116$, (h) ${^{248}}\mathrm{Cm}({^{48}}\mathrm{Ca},xn){^{296-x}}116$, (i) ${^{249}}\mathrm{Bk}({^{48}}\mathrm{Ca},xn){^{297-x}}117$ and (j) ${^{249}}\mathrm{Cf}({^{48}}\mathrm{Ca},xn){^{297-x}}118$. Lines show the calculated data in different neutron channels, summation of all channels, and dots show the experimental data.}
\end{figure*}

\begin{table}
\caption{\label{tab:table3} Combination, calculated CN excitation energy, the center of mass energy, and the maximum ER cross section to synthesize SHN with $Z=121$.}
\renewcommand{\arraystretch}{1.5}
\begin{ruledtabular}
\begin{tabular}{lcr}
Combination&$E_{CN}^\ast[E_{\mathrm{c.m.}}](\mathrm{MeV})$ &$\sigma_{ER}(\mathrm{fb})$\\ \hline
${^{252}}\mathrm{Es}({^{50}}\mathrm{Ti},3n){^{299}}121$ & 32.54[230] & $\sigma_{3n}=24.5$\\
${^{254}}\mathrm{Es}({^{50}}\mathrm{Ti},3n){^{301}}121$ & 33.13[229] & $\sigma_{3n}=11.8$\\
${^{248}}\mathrm{Cf}({^{50}}\mathrm{V},3n){^{295}}121$ & 36.62[238] & $\sigma_{3n}=0.4$\\
${^{248}}\mathrm{Cf}({^{51}}\mathrm{V},2n){^{297}}121$ & 33.50[238] & $\sigma_{2n}=0.3$\\
${^{249}}\mathrm{Cf}({^{49}}\mathrm{V},3n){^{295}}121$ & 38.94[238] & $\sigma_{3n}=0.2$\\
${^{249}}\mathrm{Cf}({^{51}}\mathrm{V},3n){^{297}}121$ & 34.61[238] & $\sigma_{3n}=1.0$\\
${^{250}}\mathrm{Cf}({^{50}}\mathrm{V},3n){^{297}}121$ & 37.52[237] & $\sigma_{3n}=0.2$\\
${^{250}}\mathrm{Cf}({^{51}}\mathrm{V},2n){^{299}}121$ & 33.88[237] & $\sigma_{2n}=0.3$\\
${^{251}}\mathrm{Cf}({^{49}}\mathrm{V},3n){^{297}}121$ & 40.81[237] & $\sigma_{3n}=0.1$\\
${^{251}}\mathrm{Cf}({^{51}}\mathrm{V},3n){^{299}}121$ & 35.55[238] & $\sigma_{3n}=1.2$\\
${^{252}}\mathrm{Cf}({^{50}}\mathrm{V},3n){^{299}}121$ & 39.35[237] & $\sigma_{3n}=0.2$\\
${^{252}}\mathrm{Cf}({^{51}}\mathrm{V},3n){^{300}}121$ & 34.06[237] & $\sigma_{3n}=0.1$\\
${^{247}}\mathrm{Bk}({^{53}}\mathrm{Cr},3n){^{297}}121$ & 33.60[245] & $\sigma_{3n}=0.3$\\
${^{247}}\mathrm{Bk}({^{54}}\mathrm{Cr},2n){^{299}}121$ & 28.71[242] & $\sigma_{2n}=0.9$\\
${^{249}}\mathrm{Bk}({^{53}}\mathrm{Cr},3n){^{299}}121$ & 34.38[244] & $\sigma_{3n}=0.3$\\
${^{249}}\mathrm{Bk}({^{54}}\mathrm{Cr},2n){^{301}}121$ & 30.22[242] & $\sigma_{2n}=0.4$\\
${^{243}}\mathrm{Cm}({^{55}}\mathrm{Mn},3n){^{295}}121$ & 32.59[252] & $\sigma_{3n}=0.1$\\
${^{245}}\mathrm{Cm}({^{55}}\mathrm{Mn},3n){^{297}}121$ & 32.78[250] & $\sigma_{3n}=0.1$\\ 
${^{247}}\mathrm{Cm}({^{55}}\mathrm{Mn},3n){^{299}}121$ & 34.61[250] & $\sigma_{3n}=0.1$\\
\end{tabular}
\end{ruledtabular}
\end{table}
			
\begin{figure*}
\includegraphics[width=59mm]{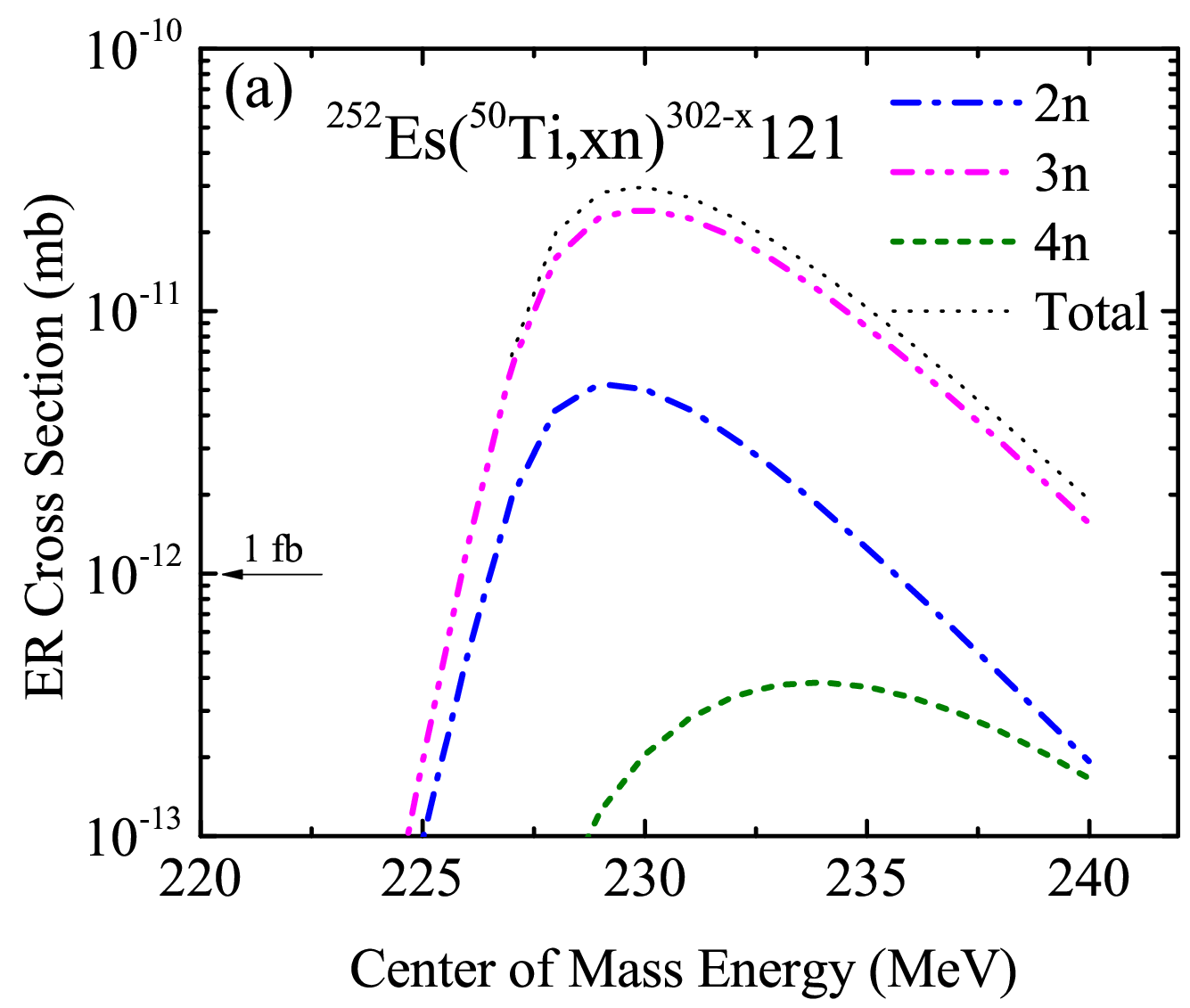}
\includegraphics[width=59mm]{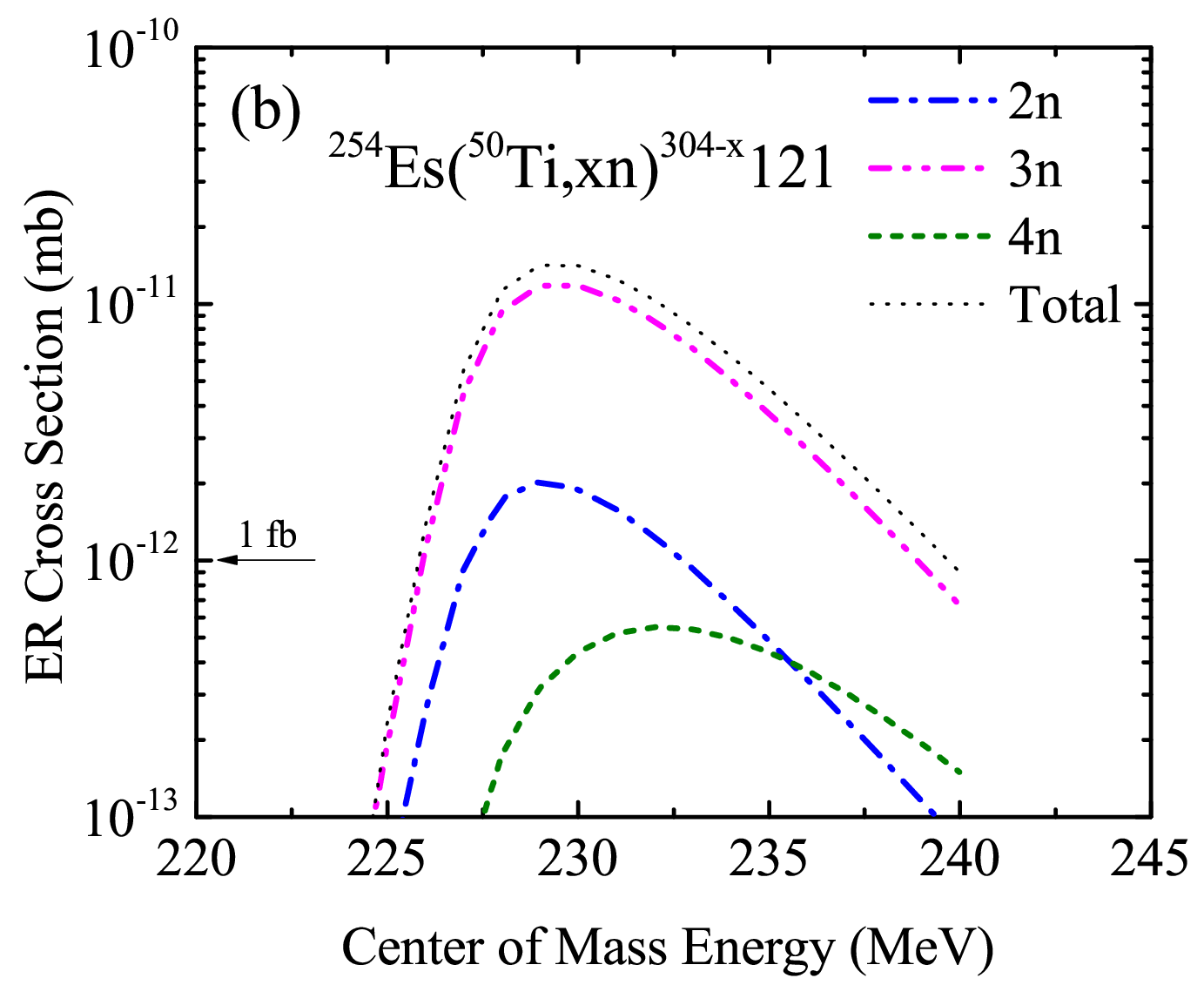}
\includegraphics[width=59mm]{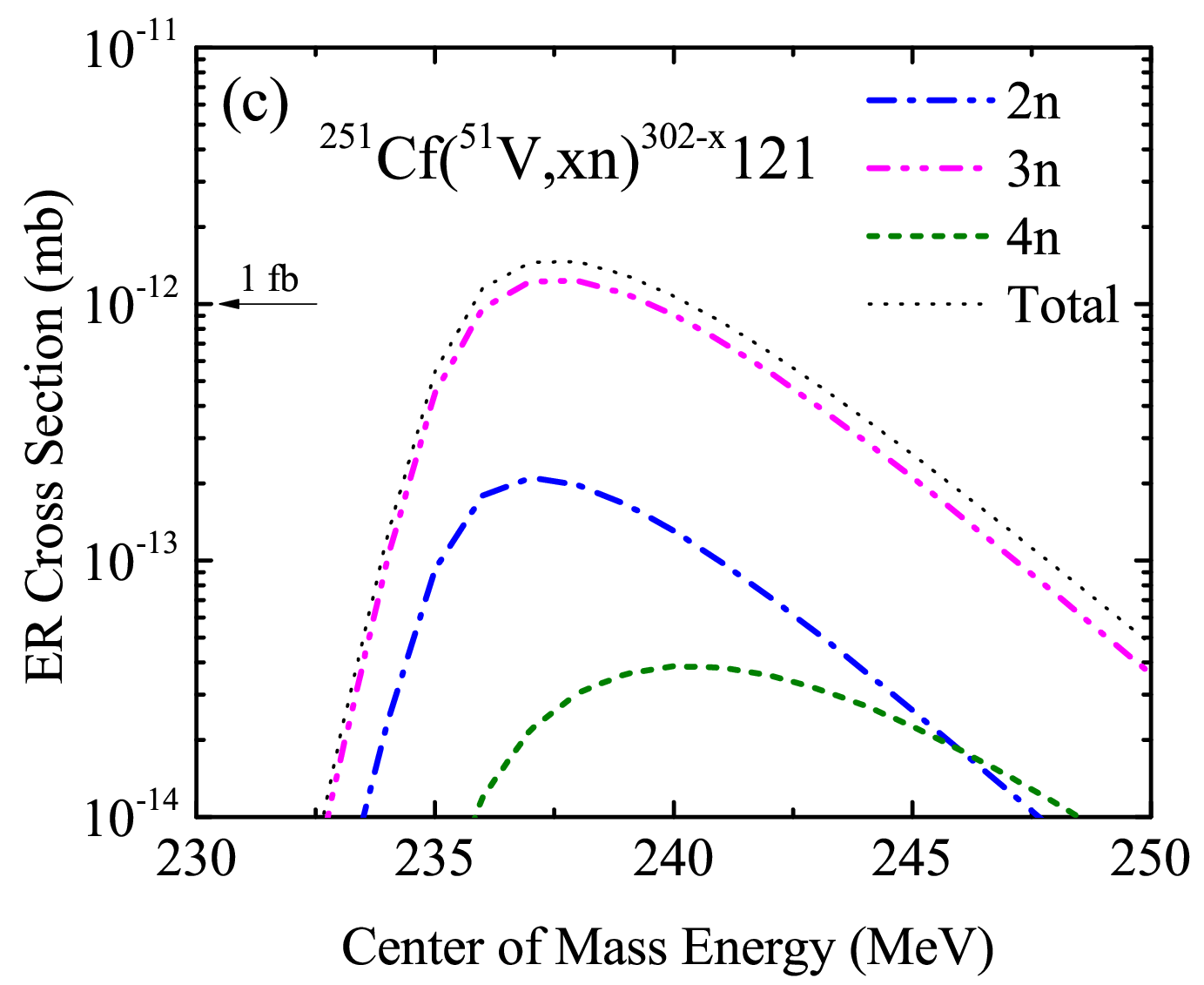}
\includegraphics[width=59mm]{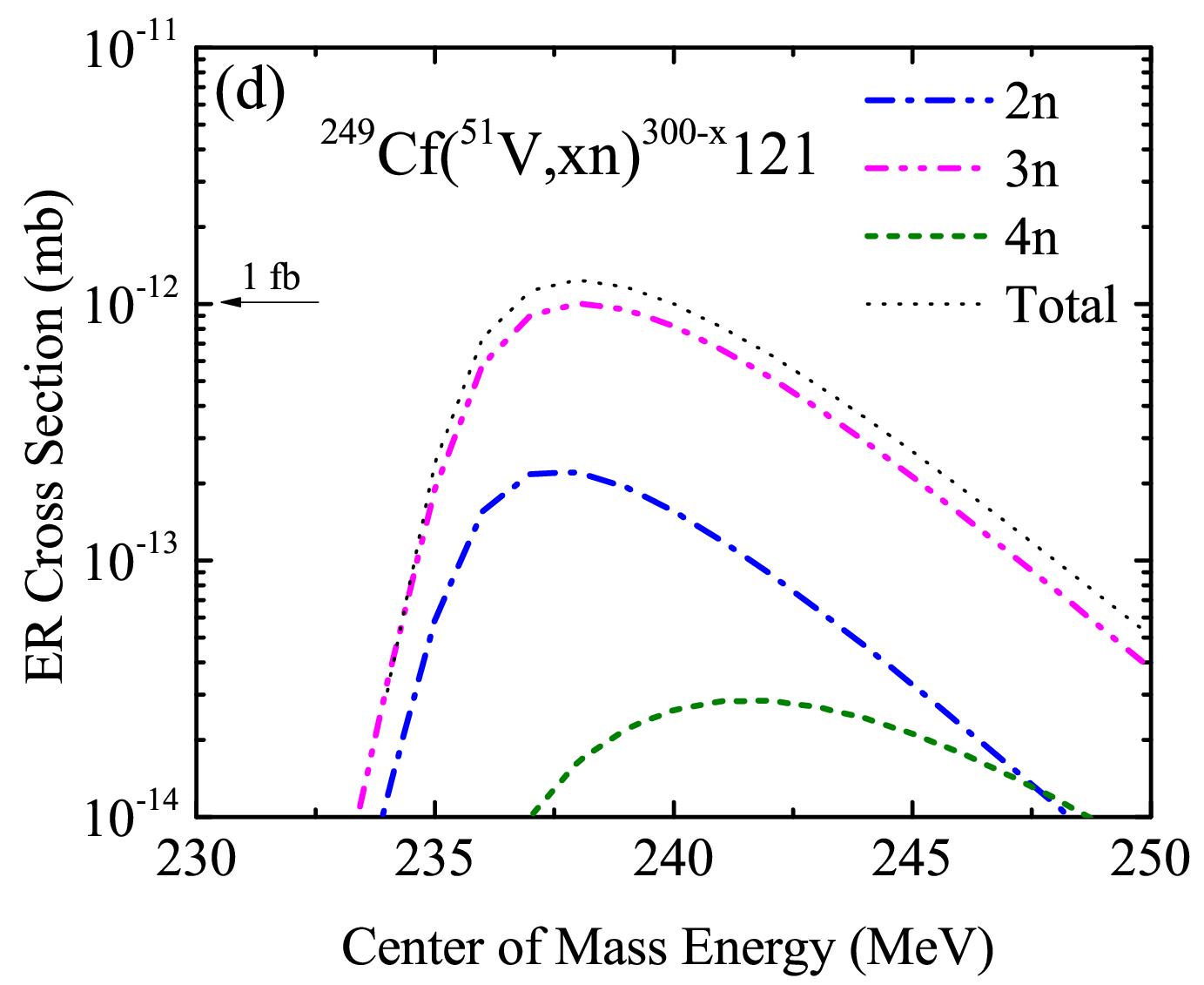}
\includegraphics[width=59mm]{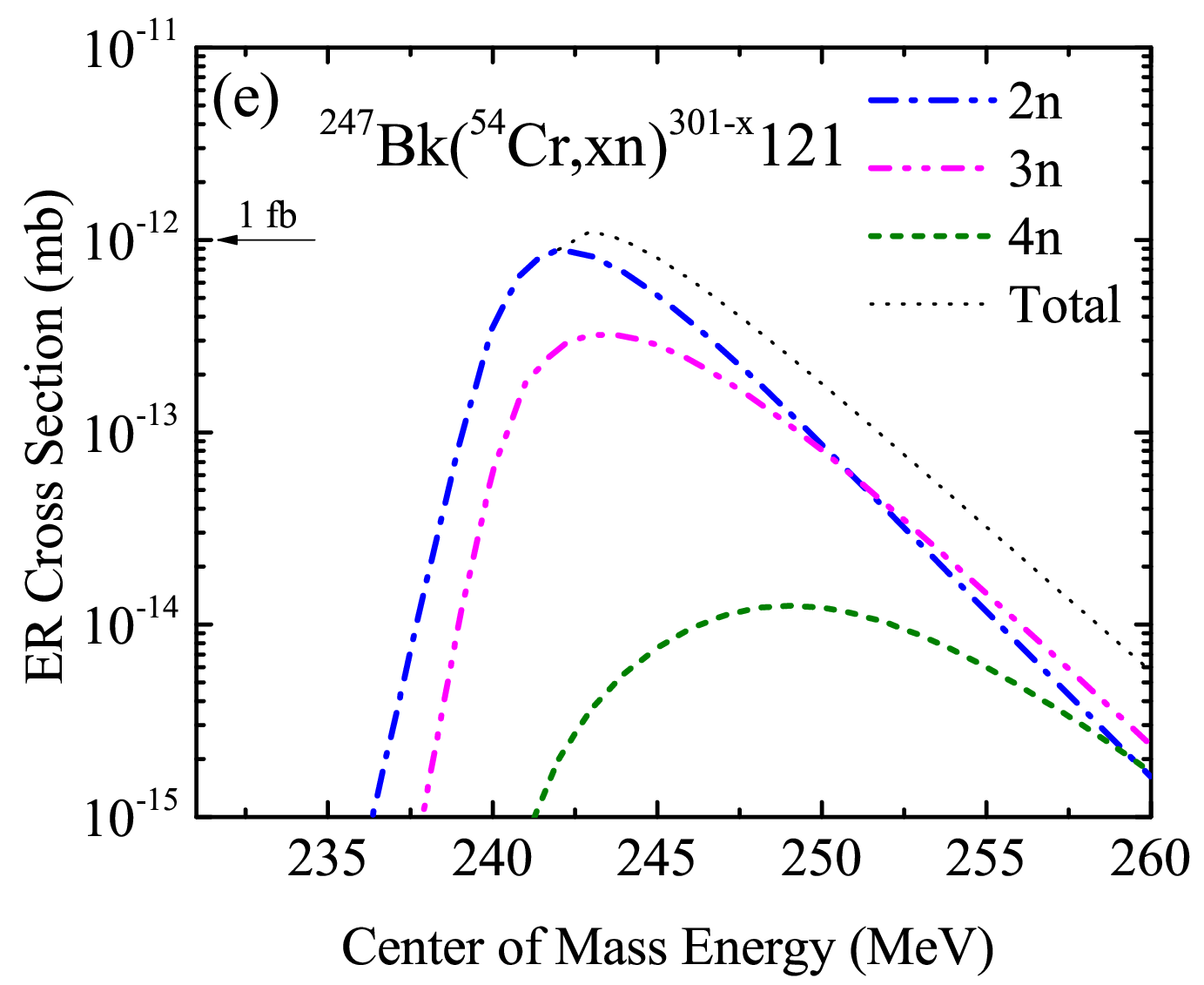}
\includegraphics[width=59mm]{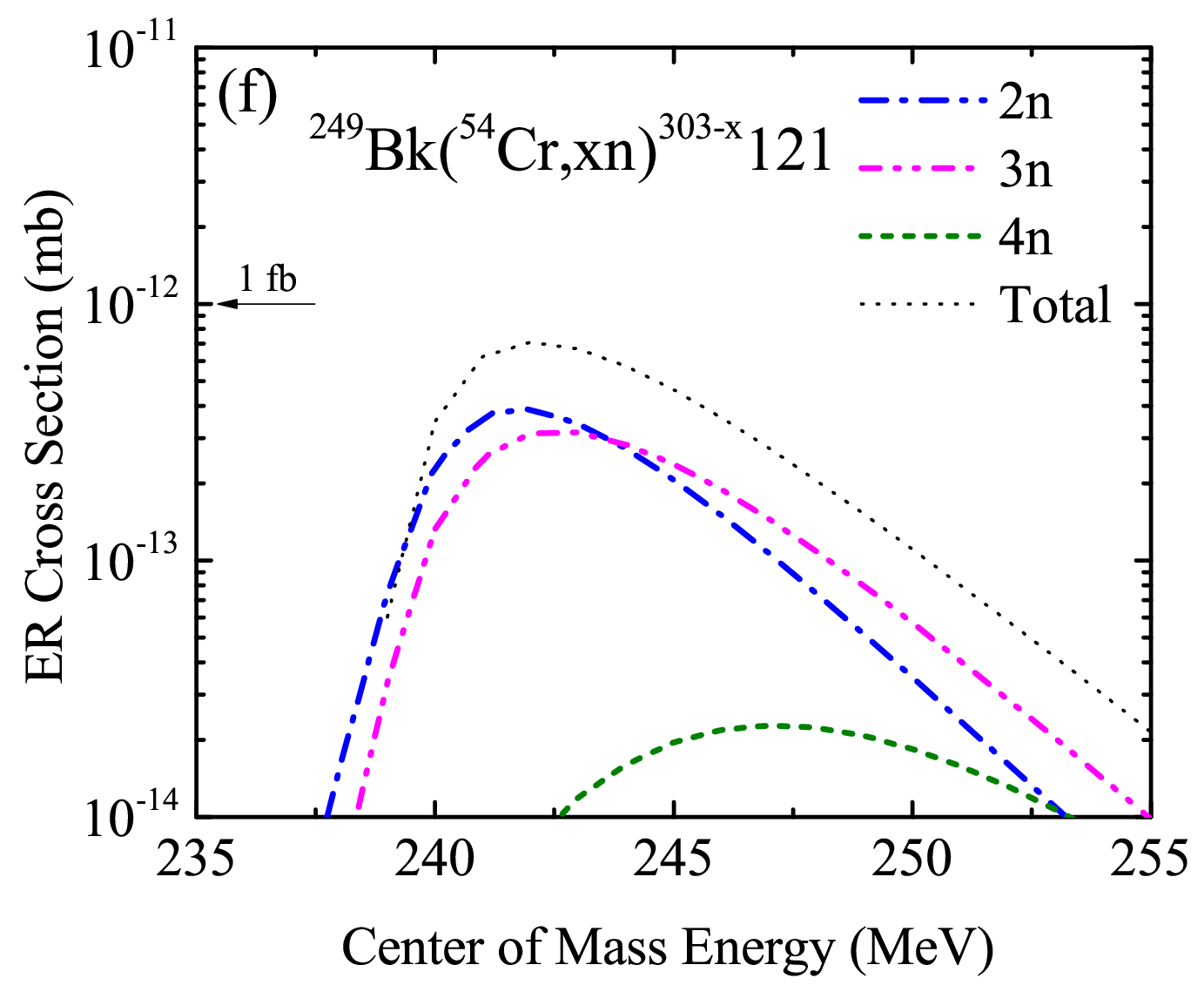}
\caption{\label{fig:fig7} The ER cross section vs the center of mass energy for combinations of (a)${^{252}}\mathrm{Es}({^{50}}\mathrm{Ti},xn){^{302-x}}121$, (b) ${^{254}}\mathrm{Es}({^{50}}\mathrm{Ti},xn){^{304-x}}121$, (c)${^{251}}\mathrm{Cf}({^{51}}\mathrm{V},xn){^{302-x}}121$, (d)${^{249}}\mathrm{Cf}({^{51}}\mathrm{V},xn){^{300-x}}121$, (e)  ${^{247}}\mathrm{Bk}({^{54}}\mathrm{Cr},xn){^{301-x}}121$, f)${^{249}}\mathrm{Bk}({^{54}}\mathrm{Cr},xn){^{303-x}}121$. Dote gray lines show the summation of all channels.}
\end{figure*}

\begin{table}%The best place to locate the table environment is directly after its first reference in text
	\caption{\label{tab:table4}
		Combinations, calculated OIE, and experimental OIE reported in Ref.~\cite{RN565}.}
	\setlength{\tabcolsep}{2pt}
	\renewcommand{\arraystretch}{1.5}
	\begin{ruledtabular}	
		\begin{tabular}{lcc}
			Combination&Calculated OIE&Experimental OIE\\ \hline
			${^{48}}\mathrm{Ca}+{^{243}}\mathrm{Am}$ & 202 & 201\\
			${^{50}}\mathrm{Ti}+{^{252}}\mathrm{Es}$ & 230 & \\
			${^{50}}\mathrm{Ti}+{^{254}}\mathrm{Es}$ & 229 & \\
			${^{51}}\mathrm{V}+{^{251}}\mathrm{Cf}$ & 238 & \\
			${^{51}}\mathrm{V}+{^{249}}\mathrm{Cf}$ & 238 & \\
			${^{54}}\mathrm{Cr}+{^{247}}\mathrm{Bk}$ & 243 & \\
			${^{54}}\mathrm{Cr}+{^{249}}\mathrm{Bk}$ & 242 &\\
		\end{tabular}
	\end{ruledtabular}
\end{table}

\begin{figure}
	\includegraphics[width=85mm]{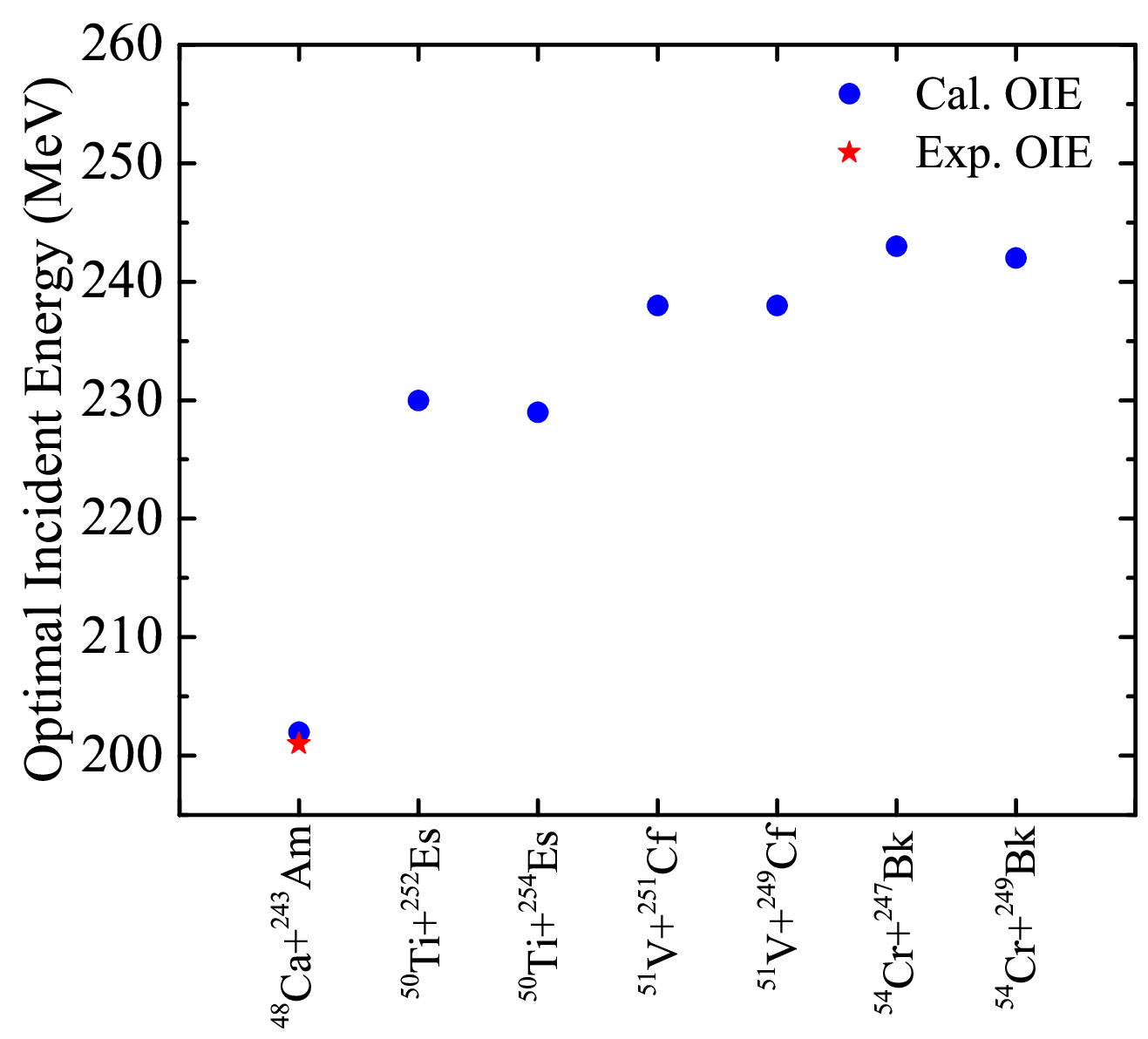}
	\caption{\label{fig:fig8} The optimal incident energy in the center of mass system for selected combinations. The experimental OIE is obtained from Ref.~\cite{RN565}.}
\end{figure}

\begin{table*}
\caption{\label{tab:table5}The comparison between the results from the current study and previously reported results employing different models.}
\renewcommand{\arraystretch}{1.5}
\begin{ruledtabular}
\begin{tabular}{ccccccc}
${^{54}}\mathrm{Cr}+{^{247}}\mathrm{Bk}$&${^{54}}\mathrm{Cr}+{^{249}}\mathrm{Bk}$&${^{50}}\mathrm{Ti}+{^{252}}\mathrm{Es}$&${^{50}}\mathrm{V}+{^{251}}\mathrm{Cf}$&${^{50}}\mathrm{Ti}+{^{254}}\mathrm{Es}$&${^{64}}\mathrm{Ni}+{^{235}}\mathrm{Np}$&Ref.\\ \hline
\shortstack{0.89($2n$) \\ 0.32($3n$)} & \shortstack{0.39($2n$),\\ 0.32($3n$),\\ 0.03($4n$)} & \shortstack{5.29($2n$),\\ 24.48($3n$),\\ 0.39($4n$)} & \shortstack{0.02($2n$),\\ 0.03($3n$)} & \shortstack{2.03($2n$),\\ 11.80($3n$),\\ 0.55($4n$)} & \shortstack{0.001($2n$),\\ 0.0005($3n$),\\ 0.00001($4n$)} & This Work\\
$>1(2n)$&&&&&&\cite{RN471}\\
&0.03803($3n$)&&&&&	\cite{RN467}\\
&&4($2n$)&&&&\cite{RN462}\\
&&&$> 1600($2n$)$&&&\cite{RN468}\\
&&&&7($4n$)&&\cite{RN461}\\
&&&&&80($3n$), 60($4n$)&\cite{RN469}\\
&\shortstack{1.45($2n$),\\ 3.28($3n$),\\ 0.45($4n$)}&\shortstack{8.62($2n$),\\ 23.1($3n$),\\ 0.33($4n$)}&&\shortstack{0.45($2n$),\\ 10.0($3n$),\\ 0.58($4n$)}&&\cite{RN470}\\
\end{tabular}
\end{ruledtabular}
\end{table*}

\section{\label{Section3}Discussion and conclusion}

\subsection{\label{Section3-1}Comparing calculated results with the experimental data}

A comparison between, the calculated results and the experimental data is presented by applying the model and theory given in sec.~\ref{Section2}. Combinations that lead to the synthesis of SHN with $Z=112-118$ were selected. These combinations included ${^{238}}\mathrm{U}({^{48}}\mathrm{Ca},xn){^{286-x}}112$, ${^{237}}\mathrm{Np}({^{48}}\mathrm{Ca},xn){^{285-x}}113$, ${^{240}}\mathrm{Pu}({^{48}}\mathrm{Ca},xn){^{288-x}}114$, ${^{242}}\mathrm{Pu}({^{48}}\mathrm{Ca},xn){^{290-x}}114$, ${^{244}}\mathrm{Pu}({^{48}}\mathrm{Ca},xn){^{292-x}}114$, ${^{243}}\mathrm{Am}({^{48}}\mathrm{Ca},xn){^{291-x}}115$, ${^{245}}\mathrm{Cm}({^{48}}\mathrm{Ca},xn){^{293-x}}116$, ${^{248}}\mathrm{Cm}({^{48}}\mathrm{Ca},xn){^{296-x}}116$, ${^{249}}\mathrm{Bk}({^{48}}\mathrm{Ca},xn){^{297-x}}117$, and ${^{249}}\mathrm{Cf}({^{48}}\mathrm{Ca},xn){^{297-x}}118$.  In Fig.~\ref{fig:fig2}, the total potential versus distance of the deformed nucleus is shown for a combination of ${^{243}}\mathrm{Am}({^{48}}\mathrm{Ca},xn){^{291-x}}115$. As shown in this figure the spherical projectile ${^{48}}\mathrm{Ca}$ collides with the oblate target ${^{243}}\mathrm{Am}$ with different target angles. The potential barrier and its position change by varying angles of the target. In Table~\ref{tab:table1} parameters from the total potential that are significant to calculate the capture cross section are shown. These parameters included the potential barrier, their positions, and inverted harmonic oscillator potential. The capture cross section differs from the fusion cross section in a heavy composite colliding system. The capture cross section determines the complete capturing projectile by the target, but not necessarily all captured nuclei will fuse. On the other hand, $\sigma_\mathrm{fusion}=\sigma_\mathrm{capture}.P_{CN}$  where $P_{CN}$ is the fusion probability, needs to be calculated precisely. Figure~\ref{fig:fig3} shows the capture cross section versus the center of mass energy for combinations of ${^{48}}\mathrm{Ca}+{^{238}}\mathrm{U}$ and ${^{48}}\mathrm{Ca}+{^{244}}\mathrm{Pu}$, along with experimental results in Refs. \cite{RN559,RN560,RN561}. The rate of increasing capture cross section up to around fusion potential is considered high; after that, it does not increase much. Figure~\ref{fig:fig4} shows the fusion probability versus the center of mass energy for the same combinations. The behavior of the fusion probability figure is similar to that of the capture cross section figure. Figure~\ref{fig:fig5} shows the survival probability versus the center of mass energy for combinations of ${^{48}}\mathrm{Ca}+{^{238}}\mathrm{U}$  and ${^{48}}\mathrm{Ca}+{^{244}}\mathrm{Pu}$. In this figure, the survival probability decreases when the center of mass energy is increased from the optimal energy. The physical reason for this is the increase in the compound nucleus's excitation energy, which causes increased fission probability. Table~\ref{tab:table2} presents the results of the calculated CN excitation energy, the center of mass energy, ER cross section, and experimental data, indicating a good agreement between the experimental data and obtained results. To synthesize elements beyond oganeson, the optimal incident energy (OIE) is an important parameter that should be extracted. This value is related to the energy obtained from the summation of all neutron channels \cite{RN562}. Figure~\ref{fig:fig6} shows the ER cross section against the center of mass energy for different neutron channels. Each figure shows the calculated ER cross section in different neutron channels, summation of channels, and experimental ER cross section obtained in Refs. \cite{RN472,RN473,RN474,RN475,RN476,RN477,RN478,RN479,RN480,RN481,RN482,RN483,RN484,RN485,RN486,RN487}

\subsection{\label{Section3-2}Probability synthesis of SHN with $\textbf{Z=121}$}

Employing the obtained model in Sec.~\ref{Section3-1}, the synthesis probability of SHN with $Z=121$ was investigated for several combinations. To synthesize superheavy nuclei beyond $Z=118$, targets heavier than californium $\mathrm{(Cf)}$ or projectiles heavier than $\mathrm{Ca}$ should be employed. All available combinations were selected to synthesize nuclei with $Z=121$. The projectiles of these combinations are stable nuclei or have a half-life of more than 300 days. In addition, the actinide targets already employed in the synthesis of SHN with $Z=113,115-118$ or targets with high half-life were used. The half-life of selected targets is more than 300 days, so the combinations of projectile and target can be confidently assumed to be realistic. Table~\ref{tab:table3} shows selected combinations, CN excitation energies, the system center of mass energies, and calculated ER cross section. From Table~\ref{tab:table3}, one can note that the best combinations to synthesize SHN with $Z=121$ are ${^{252}}\mathrm{Es}({^{50}}\mathrm{Ti},3n){^{299}}121$, with the maximum ER cross section $\sigma_{3n}=24.5~\mathrm{fb}$ at the excitation energy $E_{CN}^\ast=32.54~\mathrm{MeV}$, ${^{254}}\mathrm{Es}({^{50}}\mathrm{Ti},3n){^{301}}121$, with the maximum ER cross section $\sigma_{3n}=11.8~\mathrm{fb}$ at the excitation energy $E_{CN}^\ast=33.13~\mathrm{MeV}$, ${^{251}}\mathrm{Cf}({^{51}}\mathrm{V},3n){^{299}}121$, with the maximum ER cross section $\sigma_{3n}=1.2~\mathrm{fb}$ at the excitation energy $E_{CN}^\ast=35.55~\mathrm{MeV}, {^{249}}\mathrm{Cf}({^{51}}\mathrm{V},3n){^{297}}121$, with the maximum ER cross section $\sigma_{3n}=1.0~\mathrm{fb}$ at the excitation energy $E_{CN}^\ast=34.61~\mathrm{MeV}$, and ${^{247}}\mathrm{Bk}({^{54}}\mathrm{Cr},2n){^{299}}121$, with the maximum ER cross section $\sigma_{2n}=0.9~\mathrm{fb}$ at the excitation energy $E_{CN}^\ast=28.71 ~\mathrm{MeV}$. The ER cross section versus the CN excitation energy in two, three, and four neutron channels for some selected combinations, are shown in Fig.~\ref{fig:fig7}. In this figure, the calculated optimal incident energies calculated via the summation of all neutron channels are shown as doted gray lines. Table~\ref{tab:table4} shows the calculated optimal incident energy values. In Fig.~\ref{fig:fig8}, the optimal incident energies for the selected combinations are shown by comparing experimental data for ${^{48}}\mathrm{Ca}+{^{243}}\mathrm{Am}$ combination. The obtained ER cross sections (in femtobarn) from this work, compared with some common combinations previously reported by other research teams, are summarized in Table~\ref{tab:table5}. These combinations include ${^{247}}\mathrm{Bk}({^{54}}\mathrm{Cr},xn){^{301-x}}121$, ${^{249}}\mathrm{Bk}({^{54}}\mathrm{Cr},xn){^{303-x}}121$, ${^{252}}\mathrm{Es}({^{50}}\mathrm{Ti},xn){^{302-x}}121$, ${^{251}}\mathrm{Cf}({^{50}}\mathrm{V},xn){^{301-x}}121$, ${^{254}}\mathrm{Es}({^{50}}\mathrm{Ti},xn){^{304-x}}121$, ${^{235}}\mathrm{Np}({^{64}}\mathrm{Ni},xn){^{299-x}}121$.

\section{\label{Section4}Summary}

This paper presents the calculated ER cross section to synthesize superheavy nuclei (SHN) by employing the empirical method. The range of nuclei under study was $Z=112-118$. Theoretical calculations are in good agreement with experimental data. This model enables calculating  the ER cross section to an unknown heavier system with $Z=121$. There exist five promising combinations to synthesize SHN with $Z=121$: (1) ${^{252}}\mathrm{Es}({^{50}}\mathrm{Ti},3n){^{299}}121$, with the maximum ER cross section, $\sigma_{3n}=24.5~\mathrm{fb}$ at the CN excitation energy $E_{CN}^\ast=32.54~\mathrm{MeV} (\mathrm{OIE}=230~\mathrm{MeV})$; (2) ${^{254}}\mathrm{Es}({^{50}}\mathrm{Ti},3n){^{301}}121$, with the maximum ER cross section, $\sigma_{3n}=11.8~\mathrm{fb}$ at the CN excitation energy $E_{CN}^\ast=33.13~\mathrm{MeV}(\mathrm{OIE}=229~\mathrm{MeV})$; (3) ${^{251}}\mathrm{Cf}({^{51}}\mathrm{V},3n){^{299}}121$, with the maximum ER cross section, $\sigma_{3n}=1.2~\mathrm{fb}$ at the CN excitation energy $E_{CN}^\ast=35.55~\mathrm{MeV}(\mathrm{OIE}=238~\mathrm{MeV})$; (4) ${^{249}}\mathrm{Cf}({^{51}}\mathrm{V},3n){^{297}}121$, with the maximum ER cross section, $\sigma_{3n}=1.0~\mathrm{fb}$ at the CN excitation energy $E_{CN}^\ast=34.61~\mathrm{MeV}(\mathrm{OIE}=238~\mathrm{MeV})$; and (5) ${^{247}}\mathrm{Bk}({^{54}}\mathrm{Cr},2n){^{299}}121$, with the maximum ER cross section, $\sigma_{2n}=0.9~\mathrm{fb}$ at the CN excitation energy $E_{CN}^\ast=28.71~\mathrm{MeV}(\mathrm{OIE}=243~\mathrm{MeV})$.

% The \nocite command causes all entries in a bibliography to be printed out
% whether or not they are actually referenced in the text. This is appropriate
% for the sample file to show the different styles of references, but authors
% most likely will not want to use it.
\nocite{*}
\bibliographystyle{apsrev4-2}
%\FloatBarrier
\bibliography{SHNZ121}% Produces the bibliography via BibTeX.

\end{document}